\documentclass[preprint,review,12pt]{elsarticle}

\usepackage{amssymb}
\usepackage{amsmath}
\usepackage{lineno}
\usepackage{booktabs}
\usepackage{multirow}
\usepackage{cleveref}
\usepackage{natbib}
\usepackage[utf8]{inputenc}
\usepackage[T1]{fontenc}
\usepackage{pgf}
\usepackage{tikz}
\usepackage{pgfplots}
\usetikzlibrary{spy}
\biboptions{sort&compress}
\usepackage{xparse}
\usepackage{amsmath}
\usepackage{environ}

\ExplSyntaxOn
\NewEnviron{bmatrixT}
 {
  \marine_transpose:V \BODY
 }

\int_new:N \l_marine_transpose_row_int
\int_new:N \l_marine_transpose_col_int
\seq_new:N \l_marine_transpose_rows_seq
\seq_new:N \l_marine_transpose_arow_seq
\prop_new:N \l_marine_transpose_matrix_prop
\tl_new:N \l_marine_transpose_last_tl
\tl_new:N \l_marine_transpose_body_tl

\cs_new_protected:Nn \marine_transpose:n
 {
  \seq_set_split:Nnn \l_marine_transpose_rows_seq { \\ } { #1 }
  \int_zero:N \l_marine_transpose_row_int
  \prop_clear:N \l_marine_transpose_matrix_prop
  \seq_map_inline:Nn \l_marine_transpose_rows_seq
   {
    \int_incr:N \l_marine_transpose_row_int
    \int_zero:N \l_marine_transpose_col_int
    \seq_set_split:Nnn \l_marine_transpose_arow_seq { & } { ##1 }
    \seq_map_inline:Nn \l_marine_transpose_arow_seq
     {
      \int_incr:N \l_marine_transpose_col_int
      \prop_put:Nxn \l_marine_transpose_matrix_prop
       {
        \int_to_arabic:n { \l_marine_transpose_row_int }
        ,
        \int_to_arabic:n { \l_marine_transpose_col_int }
       }
       { ####1 }
     }
   }
   \tl_clear:N \l_marine_transpose_body_tl
   \int_step_inline:nnnn { 1 } { 1 } { \l_marine_transpose_col_int }
    {
     \int_step_inline:nnnn { 1 } { 1 } { \l_marine_transpose_row_int }
      {
       \tl_put_right:Nx \l_marine_transpose_body_tl
        {
         \prop_item:Nn \l_marine_transpose_matrix_prop { ####1,##1 }
         \int_compare:nF { ####1 = \l_marine_transpose_row_int } { & }
        }
      }
     \tl_put_right:Nn \l_marine_transpose_body_tl { \\ }
    }
   \begin{bmatrix}
   \l_marine_transpose_body_tl
   \end{bmatrix}
 }
\cs_generate_variant:Nn \marine_transpose:n { V }
\cs_generate_variant:Nn \prop_put:Nnn { Nx }
\ExplSyntaxOff
\DeclareUnicodeCharacter{2212}{−}
\usepgfplotslibrary{groupplots,dateplot}
\usetikzlibrary{patterns,shapes.arrows}
\pgfplotsset{compat=newest}
\pgfplotsset{
  compat=1.11,
  horizontal legend/.style={
    every axis legend/.append style={
      at={(-1.2,1.1)}, anchor=south west, draw=none, font=\footnotesize,
      draw=none,
      /tikz/every even column/.append style={column sep=0.4cm},
    },
    legend cell align=left,
    legend columns=5,
  },
}
\pgfplotsset{/pgfplots/group/.cd,
    horizontal sep=2.5cm,
    vertical sep=0.5cm
}
\journal{Journal Name}

\begin{document}
\tikzset{new spy style/.style={spy scope={%
 magnification=5,
 size=1.25cm, 
 connect spies,
 every spy on node/.style={
   rectangle,
   draw,
   },
 every spy in node/.style={
   draw,
   rectangle,
   fill=gray!1,
   }
  }
 }
}
\begin{frontmatter}

    \title{High-order flux reconstruction method for the hyperbolic formulation of the incompressible Navier-Stokes equations on unstructured grids}

    \author[add1]{Mohamed M. Kamra}
    \ead{mohamed.kamra@riam.kyushu-u.ac.jp}
    \author[add2]{Jabir Al-Salami}
    % \ead{jsalami@riam.kyushu-u.ac.jp}
    \author[add1]{Changhong Hu}
    % \ead{hu@riam.kyushu-u.ac.jp}
    
    \address[add1]{Research Institute for Applied Mechanics, Kyushu University,\\ 6-1 Kasuga-koen, Kasuga City, Fukuoka, Japan, 816-0855}
    \address[add2]{Interdisciplinary Graduate School of Engineering Sciences, Kyushu University,\\ 6-1 Kasuga-koen, Kasuga City, Fukuoka, Japan, 816-0855}

    \begin{abstract}
        %% Text of abstract
      A high-order Flux reconstruction implementation of the hyperbolic formulation for the incompressible Navier-Stokes equation is presented.   
      The governing equations employ Chorin's classical artificial compressibility (AC) formulation cast in hyperbolic form.  
      Instead of splitting the second-order conservation law into two equations, one for the solution and another for the gradient, the Navier-Stokes equation is cast into a first-order hyperbolic system of equations.  
      Including the gradients in the AC iterative process results in a significant improvement in accuracy for the pressure, velocity, and its gradients.  
      Furthermore, this treatment allows for taking larger time-steps since the hyperbolic formulation eliminates the restriction due to diffusion.  
      Tests using the method of manufactured solutions show that solving the conventional form of the Navier-Stokes equation lowers the order of accuracy for gradients, while the hyperbolic method is shown to provide equal orders of accuracy for both the velocity and its gradients which may be beneficial in several applications.   
      Two- and three-dimensional benchmark tests demonstrate the superior accuracy and computational efficiency of the developed  solver in comparison to the conventional method and other published works. 
      This study shows that the developed high-order hyperbolic solver for incompressible flows is attractive due to its accuracy, stability and efficiency in solving diffusion dominated problems. 
    \end{abstract}

    \begin{keyword}
        Hyperbolic method \sep Flux reconstruction \sep Unstructured grid \sep Incompressible Navier-Stokes equations \sep Artificial compressibility method

    \end{keyword}

\end{frontmatter}

% \linenumbers

\section{Introduction}
The Finite Volume method (FVM) is the most widely used method in industrial computational fluid dynamics due to its robustness and ability to handle complicated geometries using unstructured grids. These attractive features motivated a large body of research that aimed to increase its spatial accuracy beyond the standard second order while maintaining its geometric flexibility. 

Extensions of FVM to higher orders of accuracy are often achieved through reconstruction of the state variables at the cell faces based on values at neighboring cell centers\cite{Caraeni_HOFV,Gooch_HOFV}. Reconstruction strategies commonly include polynomial reconstruction\cite{Caraeni_HOFV,BARTH_HOFV}, moving Least-Squares\cite{Felgueroso_HOFV,CUETOFELGUEROSO20074712_HOFV,NOGUEIRA20102544_HOFV}, the Moving Kriging(MK)method \cite{CHASSAING2013463_HOFV} and interpolation by means of Radial basis functions (RBF) \cite{LIU20161096_HOFV,GUO201727_HOFV}. 

Most notable among high-order FVM strategies are extensions of the popular essentially non-oscillatory (ENO)
of \citeauthor{HARTEN19973_HOFV} \cite{HARTEN19973_HOFV} and the weighted ENO (WENO) schemes of \citeauthor{LIU1994200_HOFV} \cite{LIU1994200_HOFV} to unstructured grids \cite{FARMAKIS2020112921_HOFV,ZHONG2020104453_HOFV,BALSARA2020109062_HOFV,TSOUTSANIS2019100037_HOFV,BAKHVALOV2017312_HOFV}. An issue that is common among such methods is their reliance on large computational stencils. This limits their use for practical and large-scale applications due to the computational cost incurred by the increased memory access and large partitioning halo during parallel computations \cite{GARTNER2020100611_HOFV,TSOUTSANIS2018157_HOFV,ZAGHI20142151_HOFV}.

In contrast to the aforementioned approaches, high order methods, such as the discontinuous Galerkin (DG) and spectral difference (SD) methods, can achieve high order spatial accuracy on complicated geometries using compact stencils that only involve immediate face neighbors. When combined with high-order curved elements, such methods can deliver simulations of flows in complicated geometries that are more accurate than low-order methods while using fewer degrees of freedom. Nevertheless, industrial adoption of such methods remains restricted due to their large memory footprint when implicit time stepping is used. Additionally, the lack of robustness when generating higher order curved elements for regular engineering applications remains a concern. 

%%%%% FR
The flux reconstruction method (FR) was proposed by \citeauthor{Huynh2007_FR}\cite{Huynh2007_FR} to unify the nodal DG and SD methods under a single framework. In this method, the partial differential equations are solved in their differential form, similar to the finite difference (FD) method. FR schemes maintain a compact computational stencil when explicit time-stepping is used thus making it ideal for modern General Purpose Graphical Processing Units (GPGPUs) \cite{vincent2011new_FR,CASTONGUAY2013400_FR,castonguay2012new_FR,VINCENT2015248_FR,WITHERDEN2015173_FR}.
An excellent example of such implementation is the PyFR open-source code. 
PyFR is a cross-platform framework for solving advection-diffusion equations using the FR approach on mixed unstructured grids. 
This framework allows the generation of platform portable code using a single implementation via Python and MAKO templates. 
PyFR supports backends for C/OpenMP, CUDA, OpenCL and most recently HIP. Therefore it is suitable for running on CPUs as well as GPUs.
For more details on the PyFR open-source software, the reader is referred to the following works\cite{witherden2015development_FR,WITHERDEN2015173_FR,LOPPI2018193_FR,LOPPI2019108913_FR}. 
%%%%%%%%%%%%%%%%%%%%%%%%%%%%%%%%%%%
\citeauthor{VERMEIRE2017497_FR} found that high-order methods offer better accuracy vs. cost benefits relative to standard industry tools on similar hardware\cite{VERMEIRE2017497_FR}.
An FR implementation of the incompressible Naiver-Stokes equations via the AC formulation was presented by \citeauthor{COX2016414_FR}\cite{COX2016414_FR}
and \citeauthor{LOPPI2018193_FR}\cite{LOPPI2018193_FR} who later introduced adaptive local pseudo-time stepping to improve the performance of the method while maintaining accuracy\cite{LOPPI2019108913_FR}.

% hyperbolic
Lately, Nishikawa suggested a hyperbolic method for solving steady diffusion problems\cite{nishikawa2007first_hyp} and steady advection-diffusion problems \cite{nishikawa2010first_hyp} to reconcile the inconsistency between advection and diffusion fluxes.
The idea, first proposed by \citeauthor{cattaneo1958form_hyp}\cite{cattaneo1958form_hyp} and \citeauthor{vernotte1958paradoxes_hyp}\cite{vernotte1958paradoxes_hyp}, replaces the gradients of field variables that appear in the diffusive flux with additional variables that are coupled to the original system in pseudo-time. In the approach, an advection-diffusion equation (i.e. a hyperbolic-parabolic equation) is transformed in to a system of first order hyperbolic equations with a relaxation parameter that is independent of the solution or the mesh resolution.

The method was developed in the finite-volume framework for diffusion equation\cite{nishikawa2007first_hyp, NISHIKAWA2020109358_hyp,CHAMARTHI2019243_hyp,NISHIKAWA2018121_hyp,NISHIKAWA2018102_hyp},
advection-diffusion equation\cite{nishikawa2014first_hyp,NISHIKAWA201823_hyp}  
Navier-Stokes equations \cite{nishikawa2011new_hyp,nishikawa2014firstaiaa_hyp,nishikawa2015alternative_hyp,nishikawa201715_hyp}
, and incompressible Navier-Stokes equations\cite{nishikawa2014firstaiaa_hyp,Hyung2020104434}.

Furthermore, the method was adapted to the high-order DG method by \citeauthor{mazaheri2016efficient_hyp}\cite{mazaheri2016efficient_hyp}  and \citeauthor{lou2016discontinuous_hyp}\cite{lou2016discontinuous_hyp} for advection-diffusion equation on unstructured Grids. 
The method was also applied to the reconstructed discontinuous Galerkin (rDG) by \citeauthor{LOU2018103_hyp}\cite{LOU2018103_hyp} for linear advection–diffusion equations and  \citeauthor{LI2021110058_hyp}\cite{LI2021110058_hyp} for compressible Navier-Stokes equations.

\citeauthor{LOU2020109475_hyp_FR} recently developed a hyperbolic method for advection-diffusion problems in the FR framework\cite{LOU2020109475_hyp_FR} and proved its convergence, stability and consistency features for linear advection-diffusion problems for arbitrary orders of accuracy.

An issue that arises when attempting to numerically solve the conventional Navier-Stokes equation using an explicit scheme is the severe time-step restriction in diffusion dominated problems.  
Even in advection dominated problems (i.e., high Reynolds number flows), localized high diffusion areas, either due to turbulence eddy viscosity or artificially introduced viscosity for stabilization purposes, can have a significant effect on global stability especially if they overlap with locally refined mesh areas.  
Additionally, turbulence models may significantly benefit from an increase in the accuracy and order of accuracy of the velocity gradients, which are usually lower than the primitive variables in the traditional formulation. 
In this article, the artificially compressible variant of the incompressible Navier-Stokes equations are cast in a hyperbolic form and solved using the high-order flux reconstruction method. The developed hyperbolic incompressible flow solver, hereafter referred to as HINS-FR, is implemented in the framework of PyFR. 

The paper is organized as follows. In \Cref{sec:numericalmethod}, a brief overview of the flux reconstruction approach is given followed by a description of hyperbolic formulation of the incompressible NS equations. 
Key differences are highlighted between the hyperbolic and the conventional method of solving the AC-NS equations in the context of the FR approach, hereafter denoted INS-FR.   
In \Cref{sec:results}, a series of test cases are used to study the error convergence of the developed solver and results are compared with other relevant numerical and experimental studies.  
Test cases include the method of manufactured solutions, Taylor-Couette flow and the lid driven cavity problem. Finally, a 3D test case is presented for the flow past a sphere for which the hyperbolic method was compared to results of the high-order DG method as well as a hyperbolic FV method.   
The substantial improvement in parallel performance that results from using the developed HINS-FR solver is demonstrated through a scalability study. Finally, conclusions and future work are discussed in \Cref{sec:conc}. 
%% add a summary of the paper refer to \citeauthor{LOU2020109475_hyp_FR} and \cite{Hyung2020104434}

% Recent attampts for implmenting high-order reconstruction schemes in the finite-volume method on unstructured meshe has been done.
% % \newpage
\section{Numerical Method}\label{sec:numericalmethod}

%% main textP
\subsection{Flux Reconstruction}
% please paraphrase https://www.sciencedirect.com/science/article/pii/S0010465518302248
% in section 2
Consider the following conservation-law
\begin{equation}\label{stdadvec}
  {\partial u_\alpha \over \partial t} + \nabla \cdot \textbf{f}_\alpha = \mathbf{S}_\alpha
\end{equation}
where $u_\alpha = u_\alpha(\textbf{x},t) $ is the conservative field variables, $\textbf{f}_\alpha = \textbf{f}_\alpha(u_\alpha,\nabla u_\alpha) $  is the corresponding flux and $\textbf{S}_\alpha$ denotes the source term. The subscript $\alpha$ denotes a field variable, where $0 \leq \alpha < N_v$ and $N_v$ is the number of field variables. The solution domain is divided into a set of $\mathcal{N} $ non-overlapping conforming elements of suitable types such that
\begin{equation}
  \Omega = \bigcup^{\mathcal{N}-1 }_{n=0} \Omega_n
\end{equation}
where $n$ is the element index in the element set $\Omega$.

Calculations are carried out in transformed space by mapping each element into its respective canonical element according to its type.
This is achieved by means of the iso-parametric mapping
\begin{equation}
  \mathbf{\tilde{x}} = \mathcal{M} ^{-1}_{en}(\mathbf{x})
\end{equation}
\begin{equation}
  \mathbf{x} = \mathcal{M}_{en} (\mathbf{\tilde{x}} )
\end{equation}
where the subscript $e$ denotes element type. The Jacobian matrices and determinants associated with the mapping are
\begin{equation}
  \mathbf{J}_{en}^{-1} =  J_{enij}^{-1} = {\partial  \mathcal{M} ^{-1}_{eni} \over \partial x_j} \qquad\qquad
  \mathbf{J}_{en} =  J_{enij} = {\partial  \mathcal{M} _{eni} \over \partial \tilde{x}_j}
\end{equation}

\begin{equation}
  \mathcal{I}_{en} ^{-1} = det~ \mathbf{J}^{-1}_{en} = {1 \over \mathcal{I}_{en} } \qquad\qquad
  \mathcal{I}_{en} = det \mathbf{J}_{en}.
\end{equation}
The mapped flux and gradients of the solution can be expressed as 
\begin{equation}
  \mathbf{\tilde{f}} (\mathbf{\tilde{x}}, t) = \mathcal{I}_{en} ^{-1} (\mathbf{\tilde{x}}) \mathbf{J}_{en}^{-1} (\mathcal{M}_{en} (\mathbf{\tilde{x}} )) \mathbf{f}_{en\alpha} (\mathcal{M}_{en} (\mathbf{\tilde{x}} ), t )
\end{equation}
\begin{equation}
  {\tilde{\nabla} u}_{en\alpha}(\mathbf{\tilde{x}}, t) = \mathbf{J}_{en}^T(\mathbf{\tilde{x}}) {{\nabla} u}_{en\alpha}(\mathcal{M}_{en} (\mathbf{\tilde{x}} ), t)
\end{equation}
where $\tilde{\nabla} = \partial / \partial \tilde{x}_i$. \Cref{stdadvec} can then be conveniently rewritten in terms of the divergence of the transformed flux as 
\begin{equation}
  {\partial u_{en\alpha} \over \partial t} + \mathcal{I}_{en} ^{-1}\tilde{\nabla} \cdot \mathbf{\tilde{f}}_{en\alpha} = \textbf{S}_\alpha
  \label{FREquation}
\end{equation}

A set of solution points,$ \; \tilde{\mathbf{x}}_{eq}^{u}$, where $0\leq q<N_{eq}$, are distributed within $\Omega_n$ using an appropriate distribution. 
% add citation for quad rules
In this work, points in quads are the Gauss-Legendre points and Williams-Shunn\cite{williams2014symmetric} points in triangles. In three-dimensions, points in hexahedral are the Gauss-Legendre points, Shunn-Ham\cite{shunn2012symmetric} points in tetrahedra, and Gauss-Legendre-Williams-Shunn for prisms.

 Next, a nodal basis set $l_{eq}(\tilde{\mathbf{x}})$ is constructed using $\Psi(\tilde{\mathbf{x}})_{eq}$, which is a basis set that spans a polynomial space of order $m$ such that

\begin{equation}
	\mathit{l}_{eq} (\tilde{\mathbf{x}}_{er}) = \mathcal{V}_{eqr}^{-1} \Psi_{er}(\tilde{\mathbf{x}}_{er})
\end{equation}

where $\mathcal{V}_{eqr}$ are the elements of the Vandermonde matrix. The nodal basis are required to satisfy the property $l_{ei}(\tilde{\mathbf{x}}_{ej})= \delta_{ij}$.

In addition to solution points, a set of flux points ,$ \; \tilde{\mathbf{x}}_{eq}^{f}$ where $0 \leq q<N_f$, are defined on element boundaries such that they share the same physical coordinates with the flux points of face-neighbors. 

%%%%%%%%%%%%%%%%%%%%%%%%%%%%%%%%%%%%%%
% \newpage
Solving  \Cref{FREquation} using the flux reconstruction procedure can be broken-down into the following steps:

\subparagraph{Step 1-a}
The solution at an element's flux points is found by interpolating from the solution at its solution points
\begin{equation}
    u_{ein\alpha}^f = u_{eqn\alpha}^u ~l_{eq}(\tilde{\mathbf{x}}_{ei}^{f}) \qquad \forall q \in \left[ 0 , N_{eq} \right[.
\end{equation}

This results in an approximation of the solution that is discontinuous across element boundaries.

\subparagraph{Step 1-b}
The previously obtained values are then used to find a common solution at coinciding flux points from neighboring elements 
\begin{equation}
    \mathcal{C} u_{eqn\alpha}^f =\mathcal{C}  u_{e^{'}q^{'}n^{'}\alpha}^f = \mathcal{C} (u_{eqn\alpha}^f,u_{e^{'}q^{'}n^{'}\alpha}^f)\qquad \forall q \in \left[ 0 , N_e \right[
\end{equation}
where $\mathcal{C}$ here is a scalar function, commonly simple upwinding as in the local Discontinuous Galerkin (LDG) approach, that takes the left, $ u_{eqn\alpha}^f$,  and right, $u_{e^{'}q^{'}n^{'}\alpha}^f$, solutions and returns a common value.

\subparagraph{Step 1-c}
The gradient of the solution that appears in the flux, $\textbf{f}_\alpha = \textbf{f}_\alpha(u_\alpha,\nabla u_\alpha)$, is computed by reconstructing a continuous solution using a vector correction function $\mathbf{g}_{eq}^f(\mathbf{\tilde{x}})$ that satisfies
\begin{equation}
    \mathbf{ \hat{\tilde{n}}  }_r \cdot  \mathbf{g}_{eq}^f(\mathbf{\tilde{x}}_{er} ^f) = \delta_{qr}.
\end{equation}
The transformed gradient of the continuous solution at solution points is then computed as follows 
\begin{equation} \label{continuousGrad}
	\tilde{\nabla}u_{eqn\alpha}^u = (\hat{\tilde{\mathbf{n}}}_{ei} \tilde{\nabla}) \cdot \mathbf{g}_{ei}(\mathbf{\tilde{x}}_{eq}^{u}) \left(\mathcal{C} (u^{f}_{ein\alpha}) - u^{f}_{ein\alpha} \right)  + u^{u}_{ejn\alpha} \tilde{\nabla} \mathit{l}_{ej}^{u}(\mathbf{\tilde{x}}_{eq}^{u}).
\end{equation}

Gradients are transformed to physical space using
 \begin{equation} \label{interpolatedGradient}
    {\nabla}u_{eqn\alpha}^u= \mathbf{J}_{eqn}^{-T}(\tilde{\nabla}u)_{eqn\alpha}^u
 \end{equation}
 where $\mathbf{J}_{eqn}^{-T} = \mathbf{J}_{en}^{-T} (\mathbf{\tilde{x}_{eq}} )$.

Gradients are also interpolated at the flux points in a manner similar to step 1-a:

\begin{equation}
	\nabla u_{en\alpha}^f = \nabla u_{en\alpha}^u ~l_{eq}(\tilde{\mathbf{x}}_{eq}^{f}) \qquad \forall q \in \left[ 0 , N_e \right[.
\end{equation} 
 
 \subparagraph{Step 2-a}
Using the results from the steps above, the transformed flux at the solution points is evaluated according to 
\begin{equation}
    \tilde{\mathbf{f}}_{eqn\alpha}^{u} =\mathcal{I}_{eqn}\mathbf{J}^{-1}_{eqn} \mathbf{f}(u_{eqn\alpha}^u, {\nabla}u_{eqn\alpha}^u).
    \label{eq:falpha}
\end{equation}

The normal, transformed flux at flux points is then computed using
\begin{equation}\label{normalTransFlux}
    \tilde{f}_{eqn\alpha}^{f\perp} = \mathit{l}_{ei}(\tilde{\mathbf{x}}_{eq}^{f}) \hat{\tilde{\mathbf{n}}}_{eq} \cdot \tilde{\mathbf{f}}_{ein\alpha}^{u} 
\end{equation}

\subparagraph{Step 2-b} 
Common normal inviscid fluxes are found using a suitable Riemann solver. 
\begin{equation}
    \mathcal{F}_e f_{eqn\alpha}^{f\perp}  = - \mathcal{F}_e f_{e^{'}q^{'}n^{'}\alpha}^{f\perp} = 
    \mathcal{F}_e(u_{eqn\alpha}^f,
    u_{e^{'}q^{'}n^{'}\alpha}^f,\mathbf{\hat{n}}_{eqn}^f)\qquad 
\end{equation}
In all simulations carried out in this work, local Lax-Friedrichs fluxes were used. 
\subparagraph{Step 2-c} 
Similarly, a scalar function $\mathcal{F}_v$, eg. the LDG approach, is used to find the common viscous flux

\begin{equation}
	\mathcal{F}_v f_{eqn\alpha}^{f\perp}  = - \mathcal{F}_v f_{e^{'}q^{'}n^{'}\alpha}^{f\perp} = 
	\mathcal{F}_v (u_{eqn\alpha}^f,
	u_{e^{'}q^{'}n^{'}\alpha}^f, 
	\nabla u_{eqn\alpha},
	\nabla u_{e^{'} q^{'} n^{'} \alpha}
	, \, \mathbf{\hat{n}}_{eqn}^f)\qquad 
\end{equation}

\subparagraph{Step 3} 
Finally, the total normal common flux $\mathcal{F}= \mathcal{F}_e - \mathcal{F}_v$  is transformed to standard element space  
\begin{equation}
	\mathcal{F} \tilde{f}_{eqn\alpha}^{f\perp} = \mathcal{I}_{eqn}^f \mathbf{n}_{eqn}^f\mathcal{F} {f}_{eqn\alpha}^{f\perp}
\end{equation}

    and the divergence of the continuous flux is found using a procedure that is analogous to \Cref{continuousGrad}
    \begin{equation} \label{divContFlux}
        (\tilde{\nabla} \cdot \tilde{\mathbf{f}})_{eqn\alpha}^u
        =  \tilde{\nabla} \cdot \mathbf{g}_{ei}(\mathbf{\tilde{x}}_{eq}^{u}) \left(
            \mathcal{F} {f}_{ein\alpha}^{f\perp} - {f}_{ein\alpha}^{f\perp} \right)  +  \mathbf{\tilde{f}}^{u}_{ejn\alpha} \tilde{\nabla} \mathit{l}_{ej}^{u}(\mathbf{\tilde{x}}_{eq}^{u})
    \end{equation}
    This constitutes the divergence term that appears in \Cref{FREquation} to be solved via a suitable time-marching scheme. 
    % It should be noted that the source term $\textbf{S}_\alpha$ is evaluated at the solution points at the beginning of the time-step.

% \bibliographystyle{model1-num-names}
% \bibliography{references.bib}

% \end{document}

%% main textP
\subsection{Hyperbolic Incompressible Navier-Stokes Equations}
In the artificial compressibility formulation\cite{CHORIN196712}, the steady, incompressible Navier-Stokes equations are :

\begin{equation} \label{ins_eqns}
    \begin{split}
        &{\partial p \over \partial \tau}  + \nabla \cdot ( \zeta \textbf{v} ) = 0 \\
        &{\partial \textbf{v} \over \partial \tau}  + \nabla \cdot (\textbf{v}\otimes  \textbf{v} + p - \nu \nabla \textbf{v}) =0
        \end{split}
\end{equation}
where \textbf{v} is the velocity vector, $p$ is the pressure, $\nu$ is the kinematic viscosity, $\zeta$ is the the artificial compressibility relaxation factor and $\tau$ is the pseudo-time used to drive the solution to steady state.

The hyperbolic formulation of the equation is obtained by inserting $\textbf{g} = \nabla\textbf{v}$ in 
\cref{ins_eqns} and introducing an additional equation for \textbf{g}
\begin{equation} \label{hins_eqns}
    \begin{split}
        &{\partial p \over \partial \tau}  + \nabla \cdot ( \zeta \textbf{v} ) = 0 \\
        &{\partial \textbf{v} \over \partial \tau}  + \nabla \cdot (\textbf{v}\otimes  \textbf{v} + p\textbf{I} - \nu \textbf{g}) =0 \\
        & {\partial \textbf{g} \over \partial \tau} = {1 \over T_r}(\nabla\textbf{v}  - \textbf{g} )
    \end{split}
\end{equation}
where $T_r$ is the relaxation time, given by
\begin{equation}
   T_r = L^2/\nu .
\end{equation}
The length scale $L$ is defined as $L = 1/{2\pi}$. At steady-state $\textbf{g} = \nabla\textbf{v}$ and the incompressible Navier-Stokes equations are recovered. 
% More details regarding the hyperbolic formulation is cited in the previous section and thus not mentioned here.

In three dimensions, the conservative variables for the hyperbolic incompressible Navier-Stokes (HINS) equations are
\begin{equation}
    \mathbf{u} = \begin{Bmatrix}
        p,
        v_x,
        v_y,
        v_z,
        g_{xx},
        g_{xy},
        g_{xz},
        g_{yx},
        g_{yy},
        g_{yz},
        g_{zx},
        g_{zy},
        g_{zz}
        \end{Bmatrix}^T
\end{equation}
where $v_x,~ v_y,~ v_z$ are the velocity components, $g_{ij}$ is the gradient of the $i^{th}$ velocity component in the $j^{th}$ direction 
% (e.g. $g_{xy}$  is the derivative of the velocity component $v_x$ in the $x$-direction $g_{xy} = {\partial v_x \over \partial y}$)
. 
The fluxes are given as 

\begin{equation}
    \textbf{f} = 
    \begin{Bmatrix}
       f^e_{x} - f^v_{x}\\
       f^e_{y} - f^v_{y}\\
       f^e_{z} - f^v_{z}
    \end{Bmatrix}
\end{equation}
where
\begin{equation}
    \begin{split}
    &\textbf{f}^e_{x} = 
    \begin{Bmatrix}
       \zeta v_x,
           v_x^2 + p,
           v_x v_y,
           v_x v_z,
           -{1 \over T_r}v_x,
           0,
           0,
           -{1 \over T_r}v_y,
           0,
           0,
           -{1 \over T_r}v_z,
           0,
           0,
    \end{Bmatrix}^T, \\
    &\textbf{f}^e_{y} = 
    \begin{Bmatrix}
            \zeta v_y,
            v_y v_x,
           v_y^2 + p,
           v_z v_x,
           0,
           -{1 \over T_r}v_x,
           0,
           0,
           -{1 \over T_r}v_y,
           0,
           0,
           -{1 \over T_r}v_z,
           0,
    \end{Bmatrix}^T, \\
    &\textbf{f}^e_{z} = 
    \begin{Bmatrix}
            \zeta v_z,
            v_z v_x,
            v_z v_y,
           v_z^2 + p,        
           0,
           0,
           -{1 \over T_r}v_x,
           0,
           0,
           -{1 \over T_r}v_y,
           0,
           0,
           -{1 \over T_r}v_z,
    \end{Bmatrix}^T
\end{split}
\end{equation}

\begin{equation}
    \begin{split}
    &\textbf{f}^v_{x} = 
    \begin{Bmatrix}
       0,
       \nu  g_{xx},
       \nu  g_{yx},
       \nu  g_{zx},
           0,
           0,
           0,
           0,
           0,
           0,
           0,
           0,
           0,
    \end{Bmatrix}^T, \\
    &\textbf{f}^v_{y} = 
    \begin{Bmatrix}
        0,
        \nu g_{xy},
        \nu  g_{yy},
        \nu  g_{zy},
           0,
           0,
           0,
           0,
           0,
           0,
           0,
           0,
           0,
    \end{Bmatrix}^T, \\
    &\textbf{f}^v_{z} = 
    \begin{Bmatrix}
        0,
        \nu  g_{xz},
        \nu  g_{yz},
        \nu  g_{zz},      
           0,
           0,
           0,
           0,
           0,
           0,
           0,
           0,
           0,
    \end{Bmatrix}^T
\end{split}
\end{equation} 
It should be noted that the flux, $\textbf{f}_\alpha$ in \Cref{stdadvec,eq:falpha}, is no longer directly dependent on $\nabla \mathbf{u}_\alpha$ but it is simply $\textbf{f}(\mathbf{u}_\alpha) $.
Consequently, steps 1-b, 1-c and 2-c in the flux reconstruction procedure which are used to treat viscous fluxes are no longer necessary.  

The source term is defined by 
\begin{equation}
    \mathbf{S} = {1 \over T_r}\begin{Bmatrix}
        0,
        0,
        0,
        0,
        -g_{xx},
        -g_{xy},
        -g_{xz},
        -g_{yx},
        -g_{yy},
        -g_{yz},
        -g_{zx},
        -g_{zy},
        -g_{zz},
        \end{Bmatrix}
\end{equation}

For 2D problems, \citeauthor{nishikawa2014firstaiaa_hyp}\cite{nishikawa2014firstaiaa_hyp} reported that the normal flux Jacobian of the hyperbolic system has the following eigenvalues; 

\begin{equation}
    \lambda = v_n\pm \sqrt{c_a^2 + c_v^2},~~{1 \over 2}(v_n\pm \sqrt{v_n^2 + c_v^2}),0,0,0
\end{equation} 

where $v_n = v_x n_x + v_y n_y$, $c_a = \sqrt{v_n^2 + \zeta}$ and $c_v = \sqrt{\nu/T_r}$.  
As the Reynolds Number increases, the flow becomes dominated by advection, thus the contribution of $c_v$ vanishes and the original eigenvalue structure is recovered. 
The Riemann solver is modified to include the additional wave-speed $c_v$ introduced by the hyperbolic formulation.

Unlike the INS-FR solver (\verb|ac-navier-stokes| in PyFR), only one ghost state is needed per boundary condition since LDG related boundary conditions are no longer present.
 A summary of the boundary conditions currently implemented in the HINS-FR solver and the corresponding ghost states is presented in \Cref{table:pyfr_hac_bcs}. 

% In this work, the conventional AC system existing in PyFR is denoted INS-FR while the developed hyperbolic AC method is denoted HINS-FR. 
% \begin{table}[htbp!] 
%     \begin{tabular}{@{}lllll@{}}
%     \toprule
%     \begin{tabular}[c]{@{}l@{}}Boundary\\ Name\end{tabular} & No Slip Wall & Slip Wall & Velocity Inlet & Pressure Outlet \\ \midrule
%     $p$      & $p_L$           & $p_L$                & $p_L$   & $p^b$    \\
%     $v_x$    & $2v_x^w-v_{xL}$ & $v_{xL}-2n_x v_{nL}$ & $v_x^b$ & $v_{xL}$ \\
%     $v_y$    & $2v_x^w-v_{xL}$ & $v_{yL}-2n_y v_{nL}$ & $v_y^b$ & $v_{yL}$ \\
%     $v_z$    & $2v_x^w-v_{xL}$ & $v_{zL}-2n_z v_{nL}$ & $v_z^b$ & $v_{zL}$ \\
%     $g_{xx}$ & $g_{xxL}$       & 0                    & 0       & 0        \\
%     $g_{xx}$ & $g_{xxL}$       & 0                    & 0       & 0        \\
%     $g_{xx}$ & $g_{xxL}$       & 0                    & 0       & 0        \\
%     $g_{xx}$ & $g_{xxL}$       & 0                    & 0       & 0        \\
%     $g_{xx}$ & $g_{xxL}$       & 0                    & 0       & 0        \\
%     $g_{xx}$ & $g_{xxL}$       & 0                    & 0       & 0        \\
%     $g_{xx}$ & $g_{xxL}$       & 0                    & 0       & 0        \\
%     $g_{xx}$ & $g_{xxL}$       & 0                    & 0       & 0        \\
%     $g_{xx}$ & $g_{xxL}$       & 0                    & 0       & 0        \\ \bottomrule
%     \end{tabular}
%     \end{table}

\begin{table}[htbp!] 
    \centering
    \caption{Summary of ghost states for the boundary conditions implemented in HINS-FR solver in PyFR. 
    % $g_{ij}$ is the velocity gradient tensor for the $i^{th}$ velocity in the $j^{th}$ direction.\\
     The superscript $b$ and subscript $L$ denote the desired boundary value and left state respectively}
    \label{table:pyfr_hac_bcs}
    \begin{tabular}{@{}cccc@{}}
        \toprule
                 & \multicolumn{3}{c}{Boundary Condition Type}    \\ \cmidrule(l){2-4}
        \begin{tabular}[c]{@{}c@{}}State \\ Variable\end{tabular} & No Slip Wall & Velocity Inlet & Pressure Outlet \\ \midrule
        $p$      & $p_L$           & $p_L$   & $p^b$    \\
        $v_x$    & $2v_x^b-v_{xL}$ & $v_x^b$ & $v_{xL}$ \\
        $v_y$    & $2v_y^b-v_{yL}$ & $v_y^b$ & $v_{yL}$ \\
        $v_z$    & $2v_z^b-v_{zL}$ & $v_z^b$ & $v_{zL}$ \\
        $g_{ij}$ & $g_{ijL}$       & 0       & 0        \\ \bottomrule
        \end{tabular}
    \end{table}

% \begin{table}[htbp!] 
%     \centering
%     \caption{Summary of ghost states for the boundary conditions implmeneted in HINS-FR solver (HACNavierStokes) in PyFR}
%     \label{table:pyfr_hac_bcs}
%     \begin{tabular}{@{}cccc@{}}
%         \toprule
%                  & \multicolumn{3}{c}{Boundary Condition Type}    \\ \cmidrule(l){2-4}
%         \begin{tabular}[c]{@{}c@{}}State \\ Variable\end{tabular} & No Slip Wall & Velocity Inlet & Pressure Outlet \\ \midrule
%         $p$      & $p_L$           & $p_L$   & $p^b$    \\
%         $v_x$    & $2v_x^w-v_{xL}$ & $v_x^b$ & $v_{xL}$ \\
%         $v_y$    & $2v_y^w-v_{yL}$ & $v_y^b$ & $v_{yL}$ \\
%         $v_z$    & $2v_z^w-v_{zL}$ & $v_z^b$ & $v_{zL}$ \\
%         $g_{xx}$ & $g_{xxL}$       & 0       & 0        \\
%         $g_{xy}$ & $g_{xyL}$       & 0       & 0        \\
%         $g_{xz}$ & $g_{xzL}$       & 0       & 0        \\
%         $g_{yx}$ & $g_{yxL}$       & 0       & 0        \\
%         $g_{yy}$ & $g_{yyL}$       & 0       & 0        \\
%         $g_{yz}$ & $g_{yzL}$       & 0       & 0        \\
%         $g_{zx}$ & $g_{zxL}$       & 0       & 0        \\
%         $g_{zy}$ & $g_{zyL}$       & 0       & 0        \\
%         $g_{zz}$ & $g_{zzL}$       & 0       & 0        \\ \bottomrule
%         \end{tabular}
%     \end{table}
% % \newpage
\section{Results}\label{sec:results}
This article focusses on the evaluation of the spatial accuracy and convergence of the hyperbolic method for incompressible flows using the FR approach. Only steady test cases are considered in this manuscript. 
While the implementation of the developed solver in PyFR allows for unsteady simulations via dual-time stepping, we choose to omit the implicit unsteady term to simplify the analysis and ensure that the results are not affected by the error term of the unsteady scheme.
This is done by excluding the physical stepper source term.  Consequently, the discussion is restricted to laminar flows, which also serves to highlight the effect of the hyperbolic method in handling diffusion dominated flows where the conventional AC formulation struggles. 
%  for two reasons: 
% \begin{itemize}
%     \item The first reason is to ensure that the flow remains steady for the previously outlined reasons and not transit to turbulence where the unsteady term must be considered. 
%     \item The second reason is to highlight the effect of the hyperbolic method in handling diffusion dominated flows where the viscous term has the highest contribution while conventional AC formulation struggles. 
% \end{itemize}

For all test cases considered in this work, the P-multigrid technique with Vermeire's Runge-Kutta smoother\cite{VERMEIRE2020109499_FR}  is used to accelerate the convergence of pseudo-time marching. 

The accuracy and efficiency of the current implementation are demonstrated using a series of test cases, namely the method of manufactured solutions, the Taylor-Couette flow, the driven cavity problem and the three-dimensional flow past sphere.
The method of manufactured solutions and the Taylor-Couette problem are used to evaluate the order of accuracy for the solution variables and velocity gradients since the exact solution is provided.
Since exact solutions are not available for the rest of the problems, results are compared to the corresponding data found in literature that is obtained either numerically or experimentally.
% \newpage

\subsection{Method of Manufactured Solutions Case}
\label{S:1}

In this section, we apply the method of manufactured solutions (MMS) to the incompressible Navier-Stokes equations.
The method is applied to both the conventional and hyperbolic formulations of the equations.
We use the MMS procedure outlined by \citeauthor{salari2000code} \cite{salari2000code}.
The manufactured solutions for the velocity and pressure for steady incompressible laminar flow are given as follows;
\begin{equation}\label{mms_u}
    u(x,y) = u_0  sin(x) sin( y)
\end{equation}
\begin{equation}\label{mms_v}
    v(x,y) = v_0 sin(2 x) sin(2y)
\end{equation}
\begin{equation}\label{mms_p}
    p(x,y) = P_0  cos(x) cos( y)
\end{equation}
where the constants are set as $ u_0 =  P_0 = 1$ and $v_0 = -1$. This solution was chosen because it provides challenging field features for polynomial based reconstruction.
\begin{figure}[htbp]
    \centering
    \includegraphics[width=0.3\linewidth]{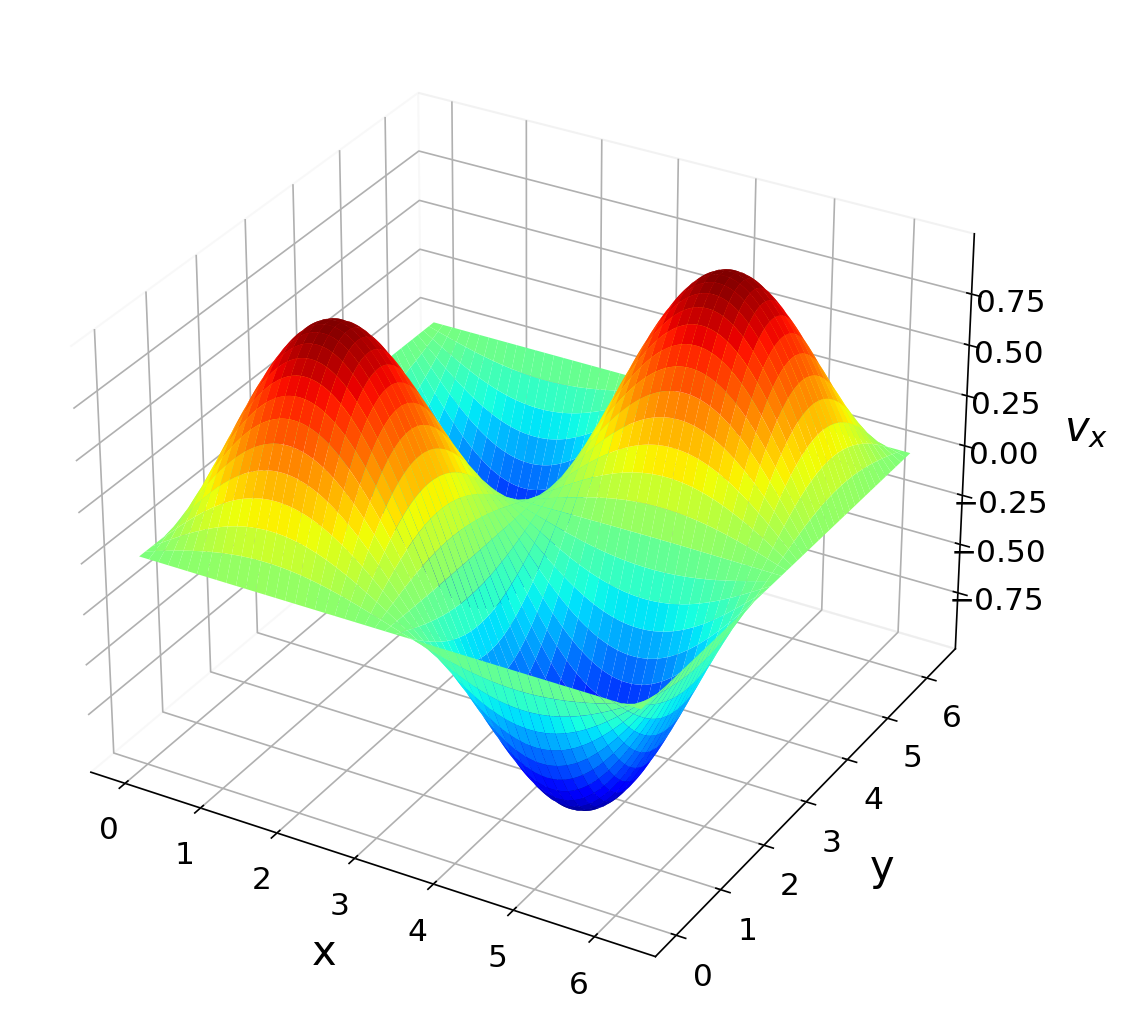}
    \includegraphics[width=0.3\linewidth]{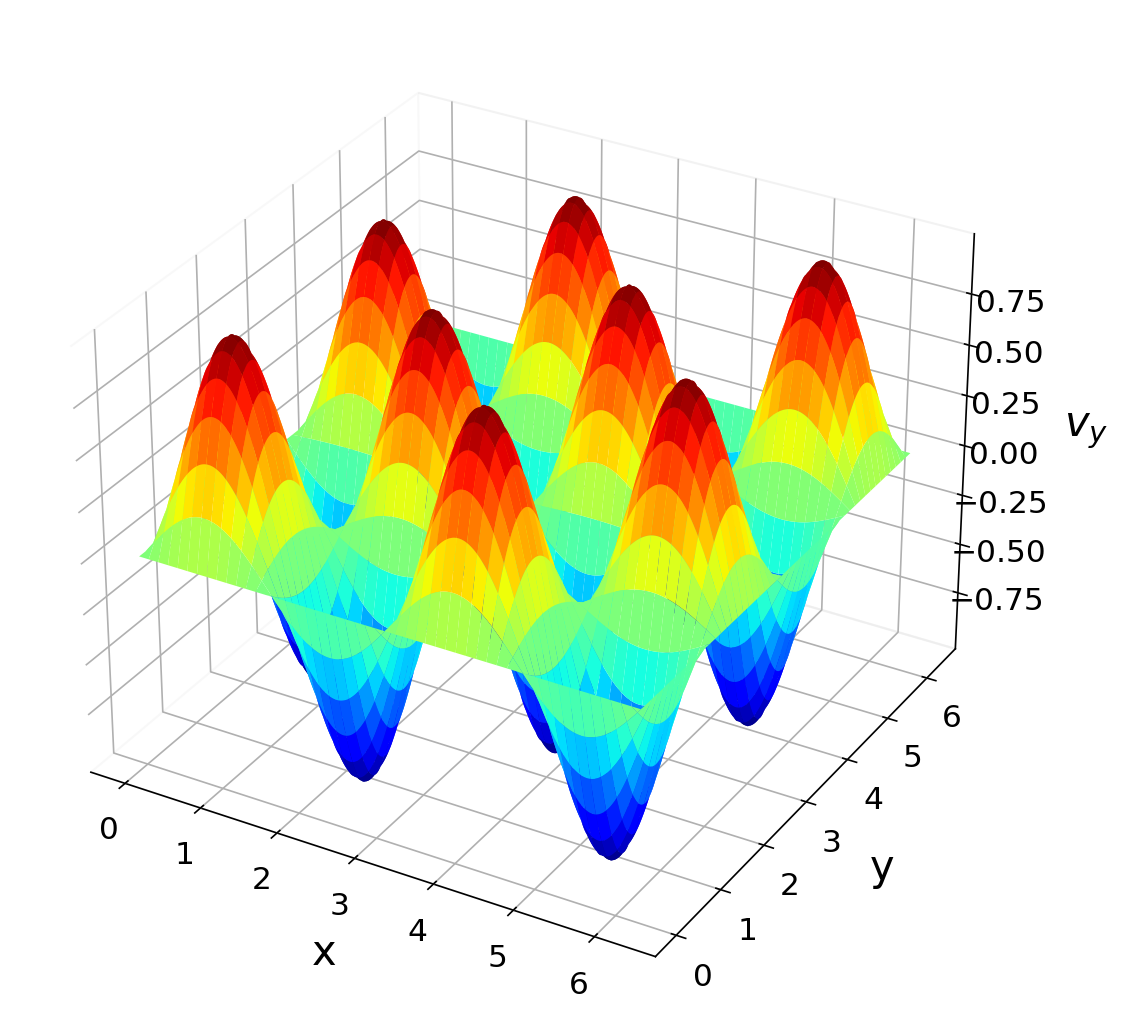}
    \includegraphics[width=0.3\linewidth]{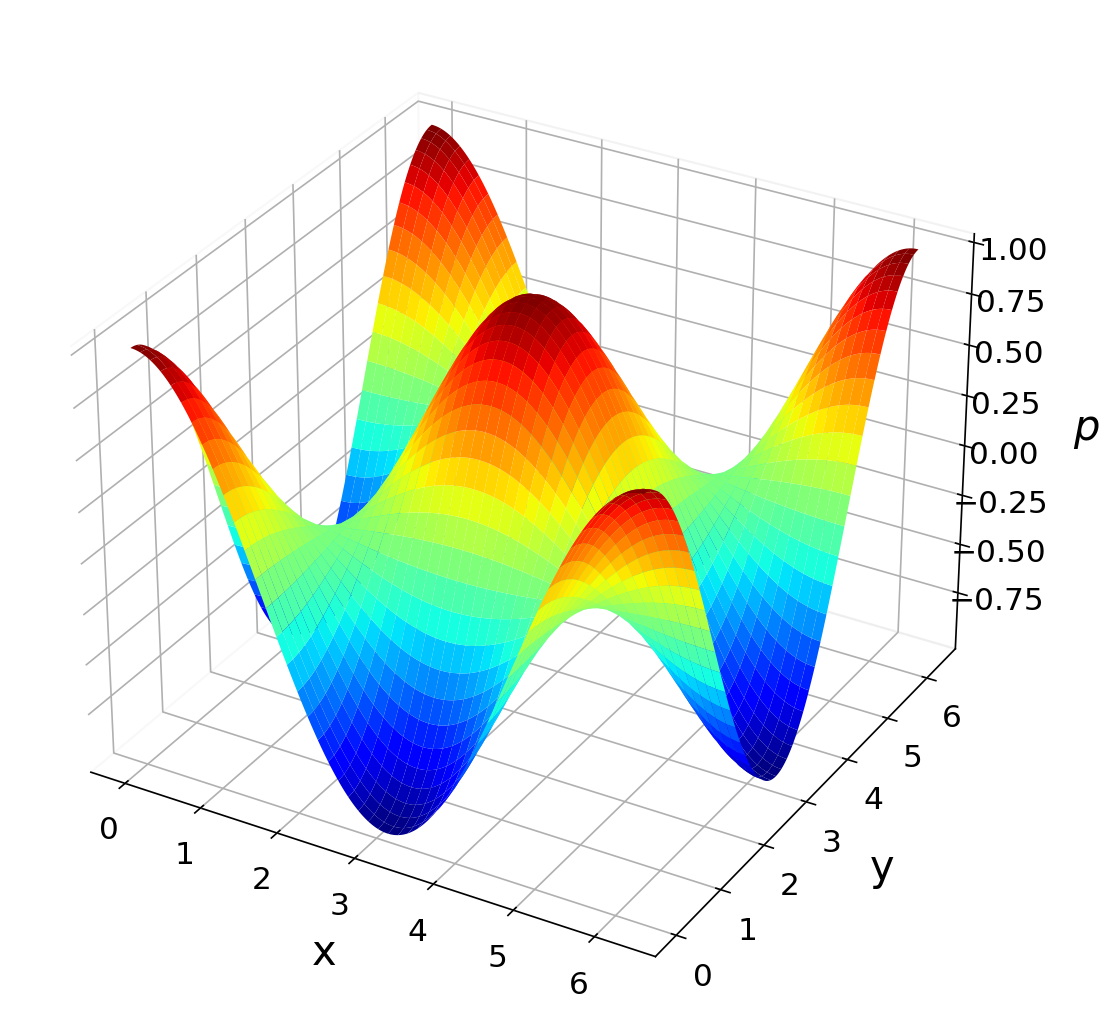}
    \caption{Manufactured solutions of the velocity components $v_x,~ v_y$ and pressure $p$ for the incompressible Navier-Stokes equations.}
    \label{figure:mansol_uvp}
\end{figure}
\Cref{figure:mansol_uvp} shows the distribution of the manufactured solution of the velocity and pressure.
%% TODO: Maybe revise and use the source term included in the report
The source terms are generated by plugging the manufactured solutions in the governing equation. Dirichlet velocity boundary conditions and Neumann pressure boundary conditions were used on all boundaries. The simulation was conducted for a kinematic viscosity $\nu$ =0.5, artificial compressibility constant $\zeta = 4 $, and a convergence tolerance of $10^{-12}$ from initial conditions equal to 1$\%$ of the manufactured solutions.

Five square domains, with a side length $d = 2\pi$, having uniform quadrilateral grids, $10\times10,~ 20\times20, 40\times40,~ 80\times80$, and $160\times160$, were used to compute the order of accuracy of the solution variable $\mathbf{u}_\alpha=(p,~v_x,~v_y)$ and the velocity gradients.
The grids are denoted by $\Omega^i$ with $\Omega^1$ being the coarsest grid and $\Omega^5$ the finest. The number of degrees of freedom  $N_{DOF}$ represents the total number of solution points in $\Omega^i$. The error measure with which the manufactured solutions was evaluated is the $L^1$-norm $\| \epsilon \|_1 $, computed using
\begin{equation}
    \| \epsilon \|_1  =  \frac{\sum_{k = 1}^{N_{DOF}}  |\epsilon| }{N_{DOF}}
\end{equation}
where $\epsilon = \widetilde{\mathbf{u}_\alpha} (x,y) - \mathbf{u}_\alpha(x,y)$, $\widetilde{\mathbf{u}_\alpha} (x,y)$  is the manufactured solution and $\mathbf{u}_\alpha(x,y)$ is the computed numerical solution, which are both evaluated at solution points.
The order of accuracy $\mathcal{P}$  is then evaluated using
\begin{equation}
    \mathcal{P} = \frac{\log{\left[\frac{\| \epsilon \|(\Omega^i)}{\| \epsilon \|(\Omega^{i+1})} \right] }} {\log{\left[\frac{h(\Omega^i)}{h(\Omega^{i+1})} \right] }}
\end{equation}
where $ h(\Omega^i) = 1/\sqrt{N_{DOF}}$ for $\Omega^i \subset \mathbb{R} ^2$. For this test case, the hyperbolic relaxation time $T_r$  for HINS-FR is set to 1/20. The choice\cite{nishikawa2007first_hyp} is based on the formula $L_r^2/\nu$ where $L_r$ taken as $1/2\pi$.

\begin{table}[htbp]
    \caption{Method of manufactured solutions problem for variable $v_x$, observed order of accuracy for HINS-FR system and INS-FR original system for $m=~3$}
    \vspace{10pt}
    \label{table:doubleSine_vx}
    \centering\begin{tabular}{lcccccc}
        \toprule
                     & \multicolumn{2}{c}{$v_x$} & \multicolumn{2}{c}{$g_{xx}$} & \multicolumn{2}{c}{$g_{xy}$}                              \\
        \midrule
        Grid         & $\| \epsilon \|_1 $             & $\mathcal{P} $                                             & $\| \epsilon \|_1 $                                       & $\mathcal{P} $ & $\| \epsilon \|_1 $& $\mathcal{P} $ \\
        \midrule
        \multicolumn{7}{l}{(a) HINS-FR} \\
        $\Omega^{1}$ &    3.15E-05 &     - &    1.32E-04 &     - &    1.74E-04 &     - \\
        $\Omega^{2}$ &    1.65E-06 &  4.26 &    7.08E-06 &  4.23 &    6.42E-06 &  4.76 \\
        $\Omega^{3}$ &    9.86E-08 &  4.06 &    4.06E-07 &  4.12 &    2.96E-07 &  4.44 \\
        $\Omega^{4}$ &    6.08E-09 &  4.02 &    2.44E-08 &  4.06 &    1.65E-08 &  4.16 \\
        $\Omega^{5}$ &    3.78E-10 &  4.01 &    1.49E-09 &  4.03 &    1.00E-09 &  4.04 \\
        \multicolumn{7}{l}{(b) INS-FR} \\
        $\Omega^{1}$ &    3.53E-05 &     - &    4.87E-04 &     - &    5.06E-04 &     - \\
        $\Omega^{2}$ &    1.91E-06 &  4.21 &    6.24E-05 &  2.97 &    5.44E-05 &  3.22 \\
        $\Omega^{3}$ &    1.12E-07 &  4.09 &    7.62E-06 &  3.03 &    6.59E-06 &  3.04 \\
        $\Omega^{4}$ &    6.99E-09 &  4.01 &    9.52E-07 &  3.00 &    8.21E-07 &  3.00 \\
        $\Omega^{5}$ &    4.39E-10 &  3.99 &    1.19E-07 &  3.00 &    1.03E-07 &  3.00 \\
        \bottomrule
    \end{tabular}
\end{table}

\begin{table}[htbp]
    \caption{Method of manufactured solutions problem for variable $v_y$, observed order of accuracy for HINS-FR system and INS-FR original system for $m=~3$}
    \vspace{10pt}
    \label{table:doubleSine_vy}
    \centering\begin{tabular}{lcccccc}
        \toprule
                     & \multicolumn{2}{c}{$v_y$} & \multicolumn{2}{c}{$g_{yx}$} & \multicolumn{2}{c}{$g_{yy}$}                              \\
        \midrule
        Grid         & $\| \epsilon \|_1 $             & $\mathcal{P} $                                             & $\| \epsilon \|_1 $                                       & $\mathcal{P} $ & $\| \epsilon \|_1 $& $\mathcal{P} $ \\
        \midrule
        \multicolumn{7}{l}{(a) HINS-FR} \\
        $\Omega^{1}$ &    1.30E-04 &     - &    9.46E-04 &     - &    7.29E-04 &     - \\
        $\Omega^{2}$ &    7.56E-06 &  4.10 &    5.96E-05 &  3.99 &    5.06E-05 &  3.85 \\
        $\Omega^{3}$ &    4.50E-07 &  4.07 &    3.82E-06 &  3.96 &    3.36E-06 &  3.91 \\
        $\Omega^{4}$ &    2.75E-08 &  4.03 &    2.44E-07 &  3.97 &    2.16E-07 &  3.96 \\
        $\Omega^{5}$ &    1.70E-09 &  4.01 &    1.55E-08 &  3.98 &    1.37E-08 &  3.98 \\
        \multicolumn{7}{l}{(b) INS-FR} \\
        $\Omega^{1}$ &    2.22E-04 &     - &    5.86E-03 &     - &    6.13E-03 &     - \\
        $\Omega^{2}$ &    1.52E-05 &  3.87 &    7.33E-04 &  3.00 &    7.97E-04 &  2.94 \\
        $\Omega^{3}$ &    9.49E-07 &  4.00 &    9.08E-05 &  3.01 &    9.93E-05 &  3.00 \\
        $\Omega^{4}$ &    5.94E-08 &  4.00 &    1.13E-05 &  3.00 &    1.24E-05 &  3.00 \\
        $\Omega^{5}$ &    3.72E-09 &  4.00 &    1.42E-06 &  3.00 &    1.54E-06 &  3.00 \\
        \bottomrule
    \end{tabular}
\end{table}

\begin{table}[htbp]
    \caption{Method of manufactured solutions problem for variable $p$, observed order of accuracy for HINS-FR system and INS-FR original system for $m=~3$}
    \vspace{10pt}
    \label{table:doubleSine_p}
    \centering\begin{tabular}{lcccc}
        \toprule
                     & \multicolumn{2}{c}{HINS-FR   } & \multicolumn{2}{c}{INS-FR }                      \\
        \midrule
        Grid         & $\| \epsilon \|_1 $                    & $\mathcal{P} $                       & $\| \epsilon \|_1 $& $\mathcal{P} $ \\
        \midrule      \\
        $\Omega^{1}$ &    4.64E-04 &     -                            &  6.32E-04 &     -     \\
        $\Omega^{2}$ &    2.27E-05 &  4.35                        & 4.71E-05 &  3.74  \\
        $\Omega^{3}$ &    1.26E-06 &  4.17                        & 4.01E-06 &  3.55  \\
        $\Omega^{4}$ &    7.57E-08 &  4.06                        & 4.30E-07 &  3.22 \\
        $\Omega^{5}$ &    4.68E-09 &  4.01                        & 5.08E-08 &  3.08  \\
        \bottomrule
    \end{tabular}
\end{table}

From the tabulated results of the velocity components, shown in \Cref{table:doubleSine_vx,table:doubleSine_vy}, it can be observed equal order of accuracy for the velocity and its gradients is obtained for the HINS-FR solver.
On the other hand, the accuracy order of the velocity gradient is one order lower than the velocity in the case of INS-FR solver.
Moreover the INS-FR method shows nearly double the absolute error values for the velocity and up to two orders of magnitude higher absolute error values for its gradients.

% A similar behavior can be observed when examining the tabulated results in \Cref{table:doubleSine_p} for pressure. The table shows that HINS-FR also recovers the design order of accuracy, which is 4 in this case, while the INS-FR does not.

% A similar behavior can be observed in the pressure results shown in \Cref{table:doubleSine_p}. More importantly, the HINS-FR solver appears to achieve equal order of accuracy for the velocity and pressure  while the INS-FR does not.
% This appears to be the case for other conventional artificial compressibility solvers\cite{MANZANERO2020109241,salari2000code,BASSI2018186666} where the pressure convergence order is consistently smaller than the velocity.

Similar behavior can be observed for the pressure when examining the results shown in \Cref{table:doubleSine_p}. The pressure error produced by the HINS-FR solver is an order of magnitude lower when compared to that of the INS-FR solver. Furthermore, both velocity and pressure error converge at the same rate using HINS-FR, while INS-FR's pressure error lags behind by up to one order of accuracy.
INS-FR's results are consistent with other conventional artificial compressibility solvers \cite{MANZANERO2020109241,salari2000code,BASSI2018186666} where the pressure convergence order is consistently smaller than that of the velocity.
% This maybe attributed to the fact that in the artificial compressibility method, the pressure evolution equation is a function of the velocity gradients. Hence, increasing the accuracy of velocity gradients leads to a significant improvement in both the order of convergence and magnitude of pressure errors.
% This maybe attributed to the fact that for the incompressible, the accuracy of the pressure is proportional to the accuracy of the divergence of velocity  (thus the velocity gradient).
%  Since the absolute errors are significantly smaller in the HINS-FR and higher order of convergence can be obtained, the pressure also has lower absolute errors and higher order of convergence.
% This maybe attributed to the stronger coupling between the pressure, velocity and velocity gradient which not only improves the accuracy of the pressure but also its error convergence.
% we notice not only that our method achieves a higher order of convergence for the pressure, but also that the absolute errors are significantly smaller

% % \newpage

%% main textP
% \subsection{Taylor-Couette with curved boundaries}
\subsection{Taylor-Couette flow}

% The current implementation of high-order flux-reconstruction for both incompressible navier stokes formulation on curved boundaries is verified using the Taylor-Couette flow.

% The current implementation is verified for problems curved boundary elements
In this section, we simulate the Taylor-Couette flow\cite{taylor1923viii} to showcase the superiority of the current implementation for problems with curved boundary elements.
The flow is created due to the rotation of two infinitely long coaxial cylinders with fluid in-between them. 
If the inner cylinder with radius $R_{1}$ is rotating at constant angular velocity $\omega _{1}$ and the outer cylinder with radius $R_{2}$ is rotating at constant angular velocity $\omega _{2}$, then the azimuthal velocity component $v_{\theta }$ at any angle $\theta $ is
\begin{equation}
   v_{\theta }(r)=Ar+{\frac {B}{r}},\quad A=\omega _{1}{\frac {\tilde{\omega} -\tilde{R} ^{2}}{1-\tilde{R} ^{2}}},\quad B=\omega _{1}R_{1}^{2}{\frac {1-\tilde{\omega} }{1-\tilde{R} ^{2}}}
\end{equation}

where
\[
  \tilde{\omega} ={\frac {\omega _{2}}{\omega _{1}}},\quad \tilde{R} ={\frac {R_{1}}{R_{2}}} 
  \]

  \begin{figure}[htbp]
    \centering
    \includegraphics[trim=0 0 1100 0,clip,width=0.47\columnwidth]{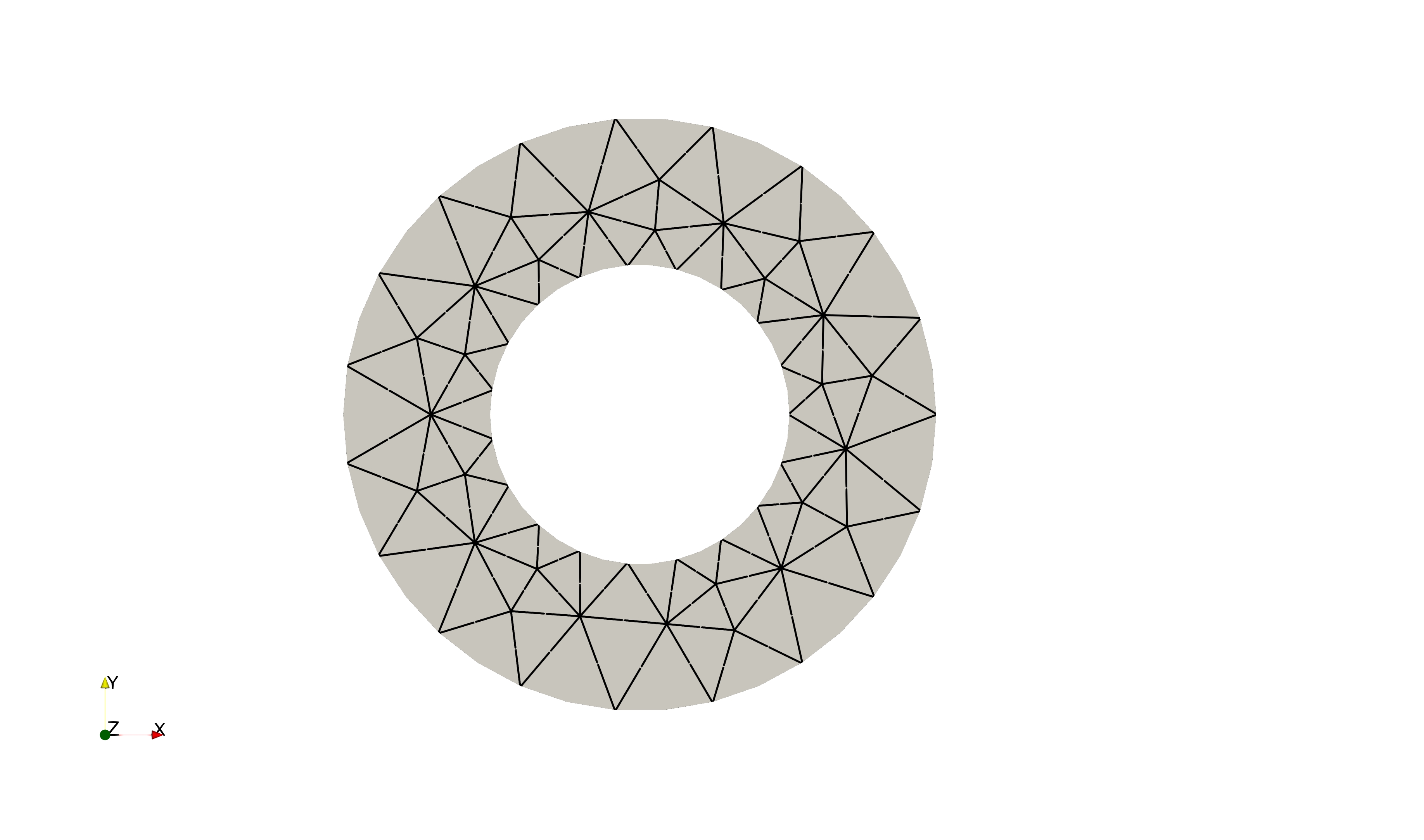}
    \includegraphics[width=0.47\columnwidth]{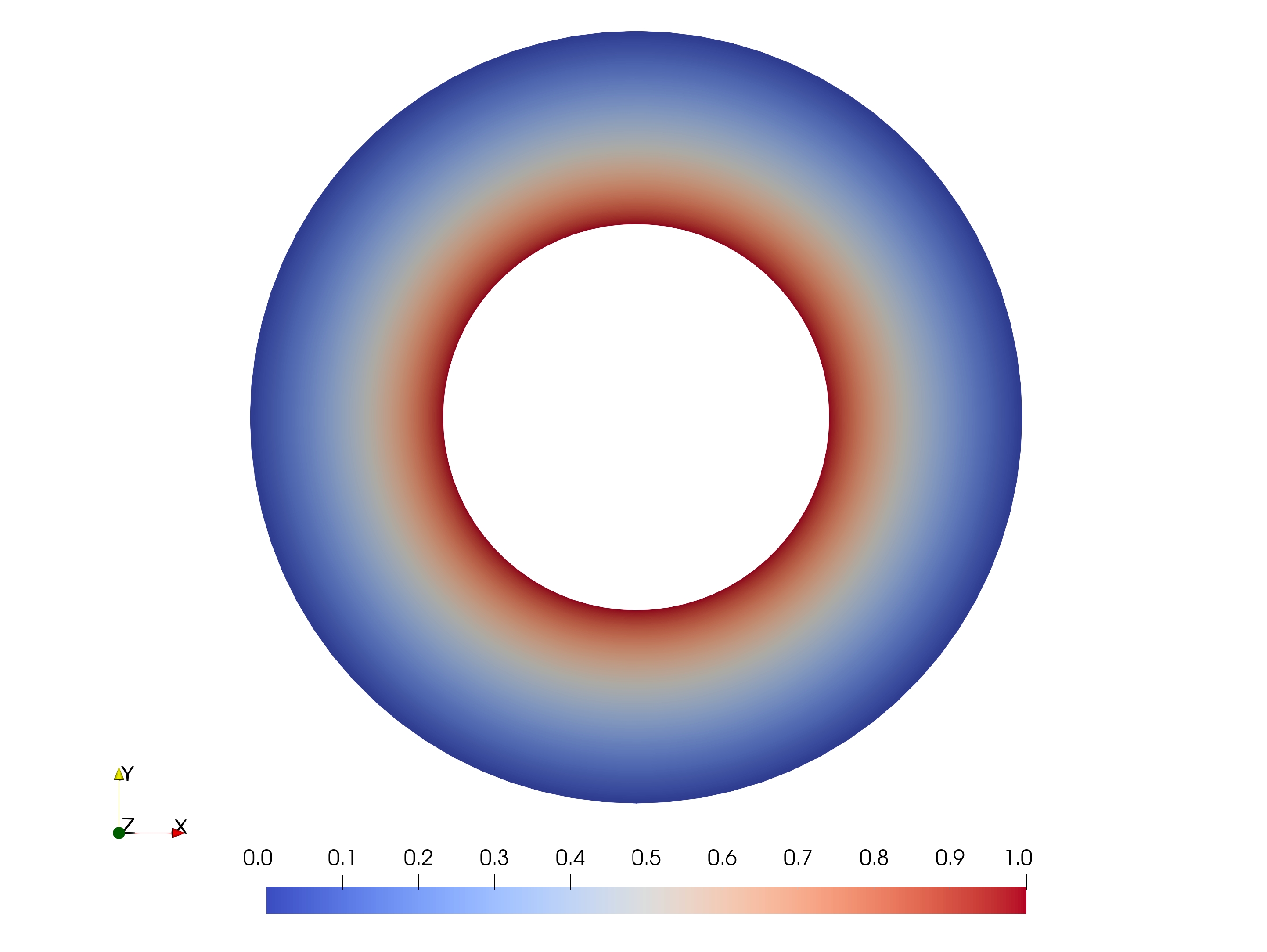}
    \caption{
      % taylor Couette Mesh (Triangular Mesh Avg Mesh spacing 0.374m )
      Incompressible Taylor-Couette flow at Re = 10 on triangular mesh:  coarsest mesh (left),  azimuthal velocity $V_\theta$ for $m=~3$ on finest mesh
      (right) }
    \label{tcouette_mesh_sol}
  \end{figure}
  In this work, the outer cylinder is stationary (i.e., $\omega_2 = 0$)  and the inner cylinder spins counter clockwise at rate $\omega_1 = 1$. The inner radius $R_1$ and other radius $R_2$ are set 1 and 2 respectively. The Reynolds number $Re ={ v_{\theta1} d_a \over \nu} $ is set to $10$, where $v_{\theta1}$ is the speed of the inner cylinder   (i.e., $v_{\theta1} = {v_\theta}_{r={R1}}  = \omega_1R_1$), $d_a$ is annulus width (i.e., $d_a = R_2 - R_1$) and $\nu$ is the kinematic viscosity.

  The order  of  accuracy  is  tested  on  a  sequence  of  four uniformly refined triangular meshes,  using  second-order  to  fourth-order  FR  schemes. 
  Quadratic curved elements are used on the boundaries. \Cref{tcouette_mesh_sol} shows the coarsest mesh of the sequence and the computed solution on the finest mesh for third-order reconstruction  scheme (i.e. $m=~ 3$ ). 
 
  % The expected azimuthal velocity contours is also shown in \Cref{tcouette_mesh_sol} (right).

% \begin{figure}[htbp]
%   \centering
%   \resizebox{0.25\columnwidth}{!}{%
%     \input{images/taylorCouetteOrderOfAccuracy_u.tex}\newline
%   }
%   \resizebox{0.25\columnwidth}{!}{%
%     \input{images/taylorCouetteOrderOfAccuracy_gxx.tex}
%   }
%   \resizebox{0.4\columnwidth}{!}{%
%     \input{images/taylorCouetteOrderOfAccuracy_gxy.tex}
%   }
%   \caption{Comparison between computed results and published literature}
%   \label{<label>}
% \end{figure}

\begin{figure}[htbp]
  \centering
  \includegraphics[trim=0 0 260 0,clip,height=0.37\columnwidth]{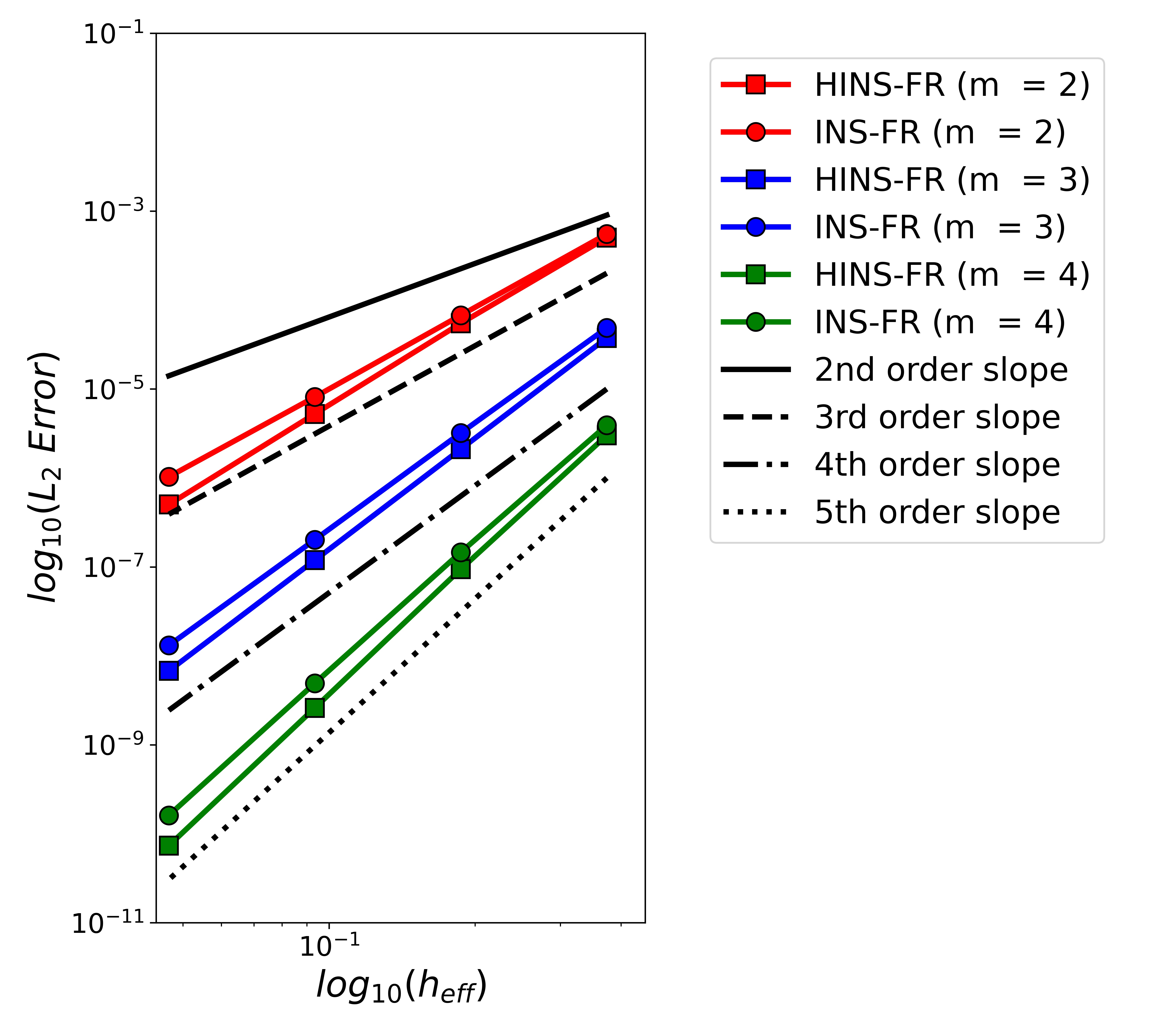}
  \includegraphics[trim=0 0 260 0,clip,height=0.37\columnwidth]{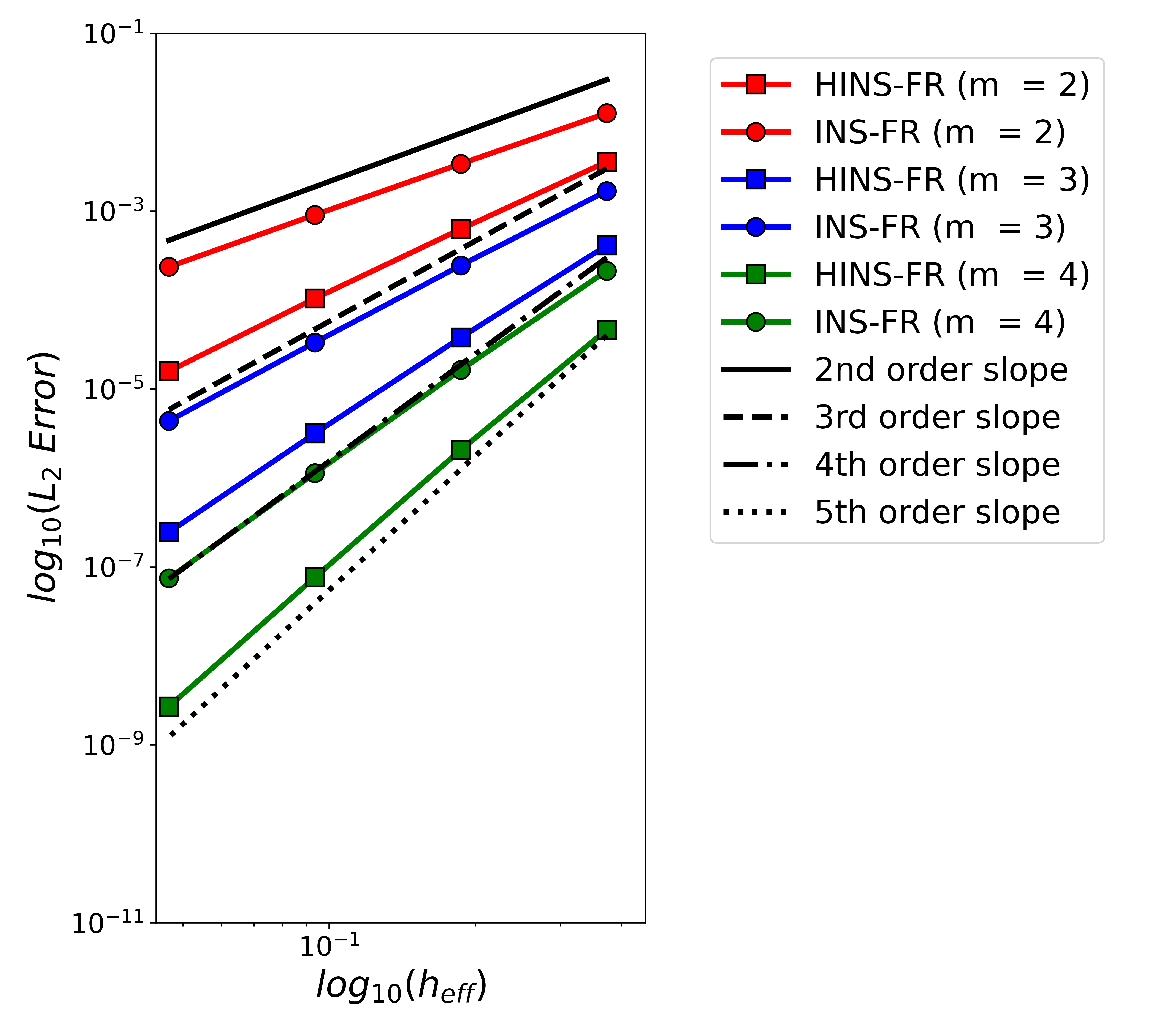}
  \includegraphics[height=0.37\columnwidth]{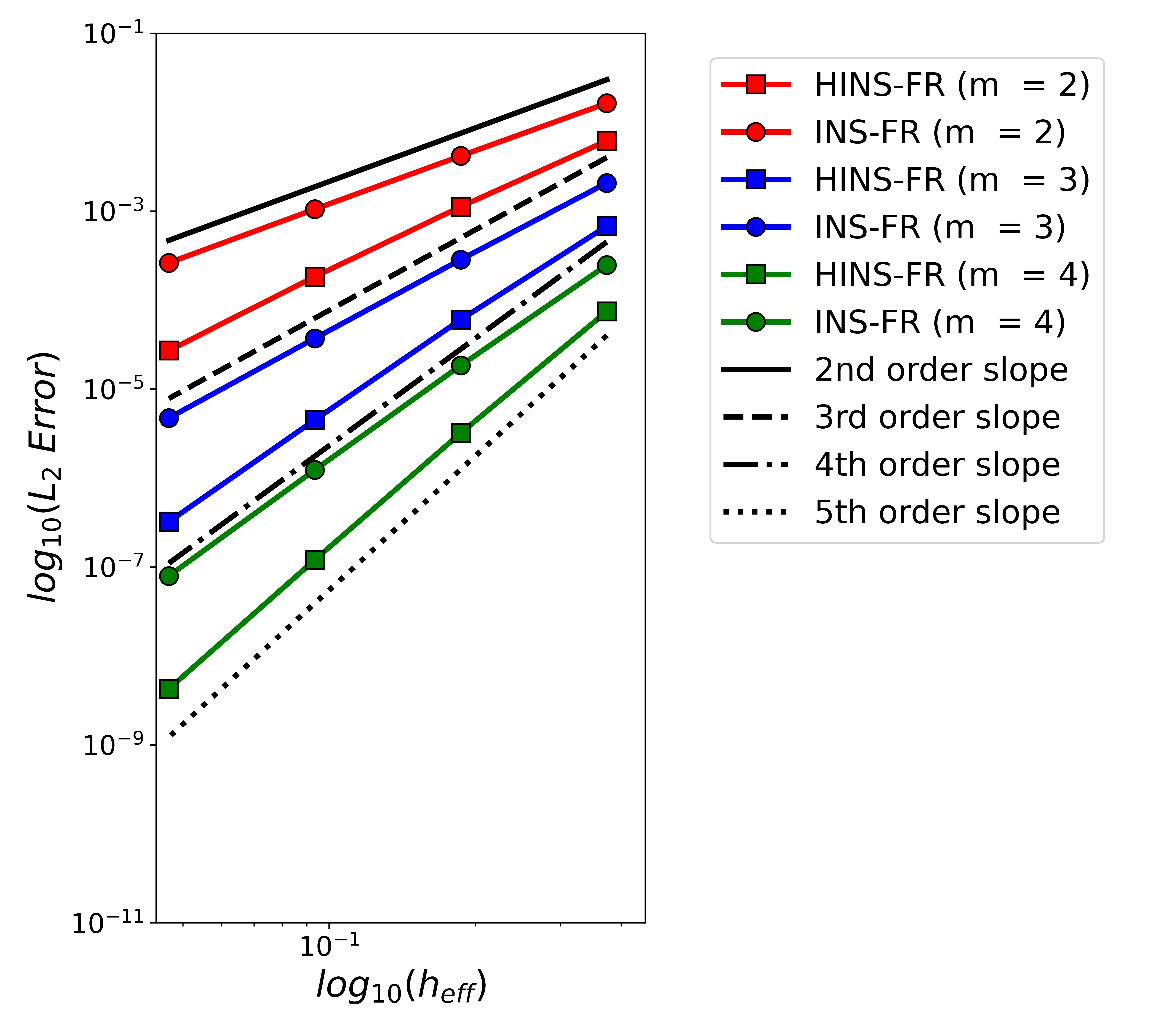}
  % \caption{Comparison of the order of accuracy computed by the  HINS-FR and INS-FR methods for the variable $u$ (left) and its
  %  gradient $\partial u \over \partial x $  (middle) and $\partial u \over \partial y $  (right) for the taylor Couette problem}
  \caption{Error convergence of or the variable $u$ (left) and its
  gradient $\partial u \over \partial x $  (middle) and $\partial u \over \partial y $  (right) for HINS-FR and INS-FR methods in the Taylor-Couette problem}
   \label{tcouette_error_order_accuracy}
\end{figure}

The reductions in $L_2$ error norm are plotted in \Cref{tcouette_error_order_accuracy} for both HINS-FR and INS-FR solvers. 
The resulting orders of accuracy are compared with expected  orders of accuracy for both the velocity and its gradient. Since the problem is symmetric only the x-velocity component $v_x$ is considered in this comparison.
The HINS-FR solver consistently produces more accurate results for the velocity.
Moreover, velocity gradient errors are significantly improved when compared to the INS-FR solver by nearly one order of magnitude.
Confirming conclusions from the previous test case, INS-FR velocity gradients converge at a rate one order below that of the HINS-FR.
% % \newpage

%% main textP
\subsection{Driven Cavity}
\begin{figure}[htbp]
    \centering
    
    \includegraphics[height=0.35\linewidth]{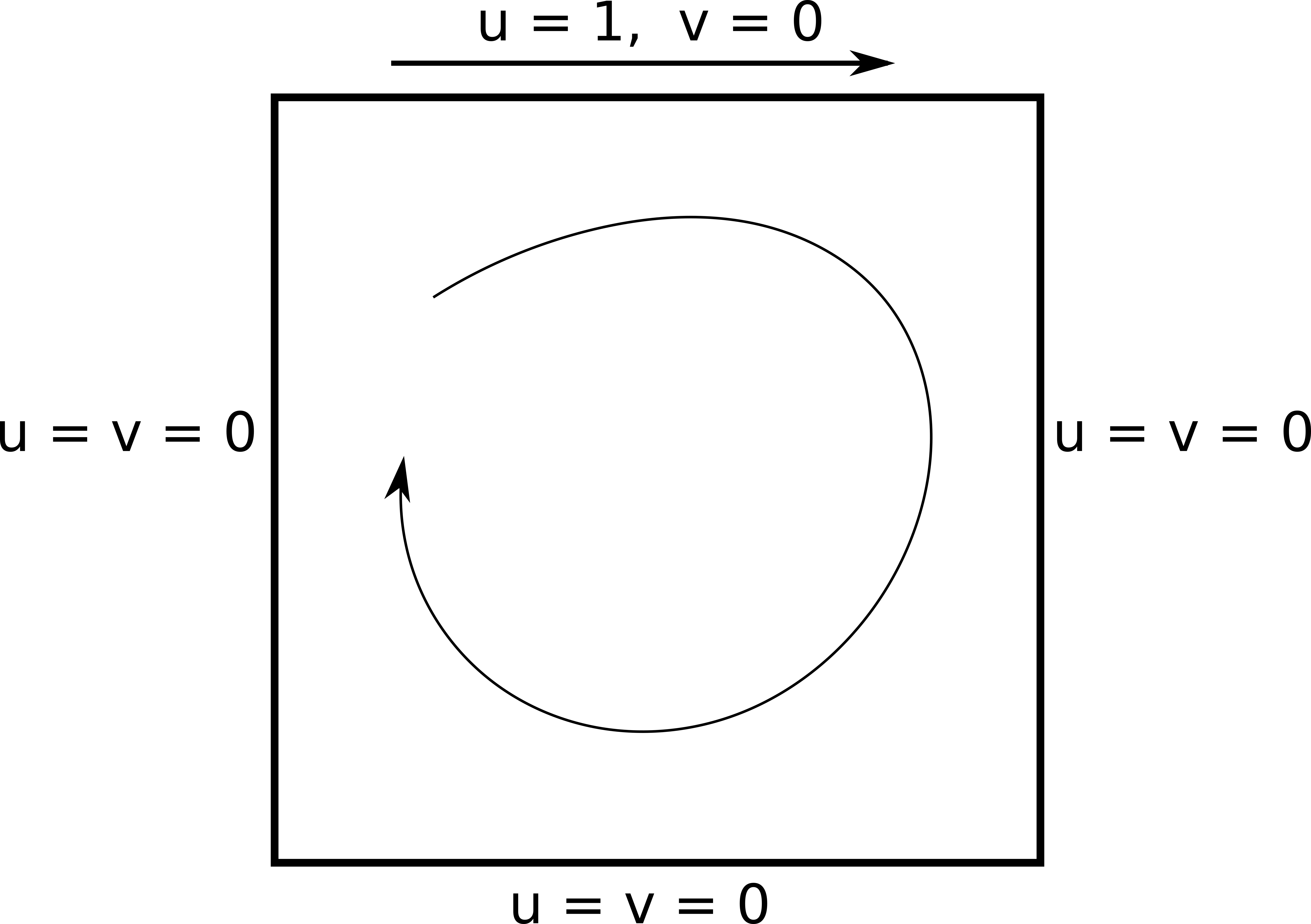}
    \quad
    \includegraphics[height=0.35\linewidth]{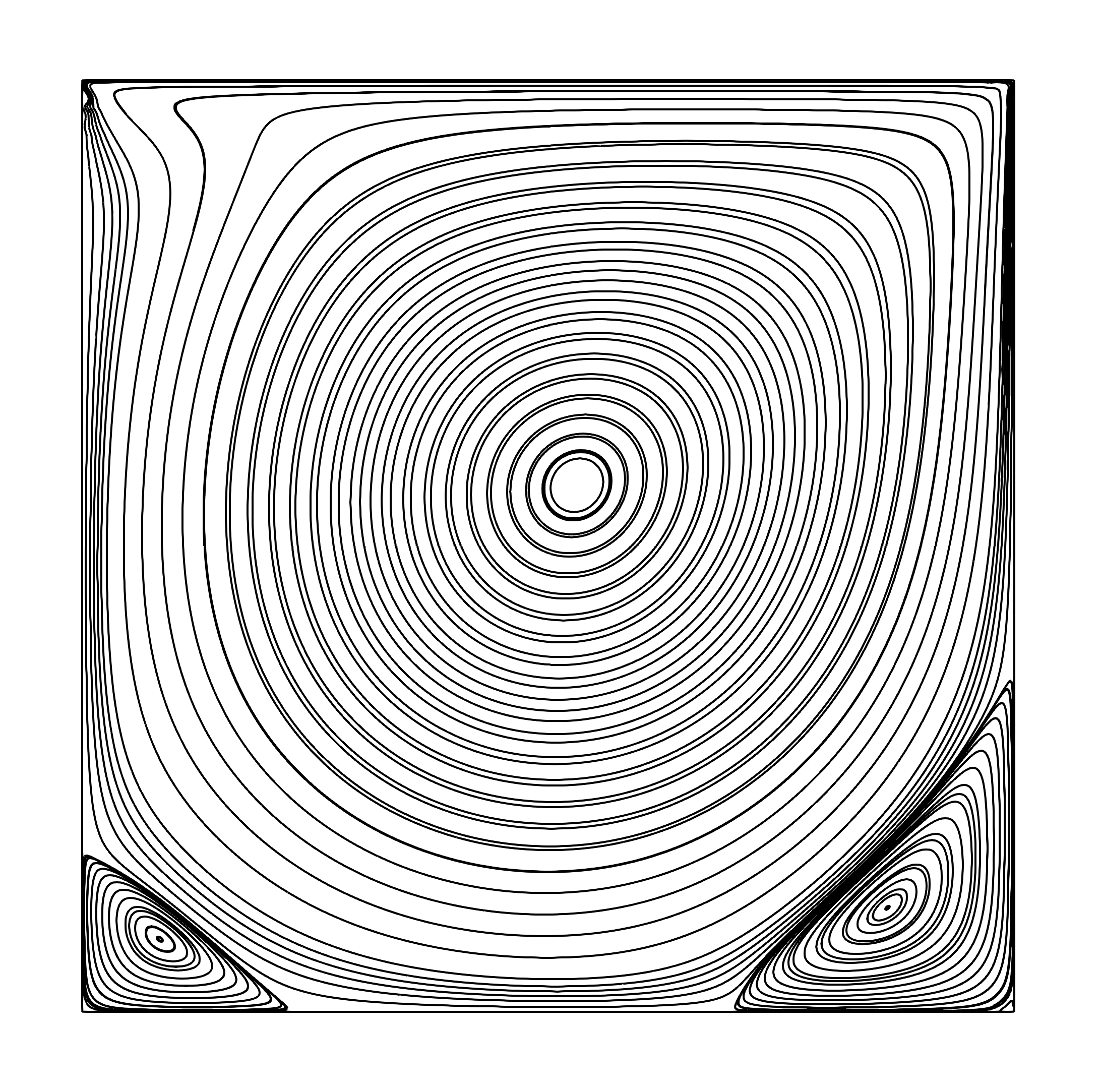}

    \caption{Driven cavity problem schematic and streamlines pattern}
    \label{figure:drivencavity_pattern}
\end{figure}

\begin{figure}[htbp!]
    \centering
    \resizebox{\columnwidth}{!}{%
    \includegraphics{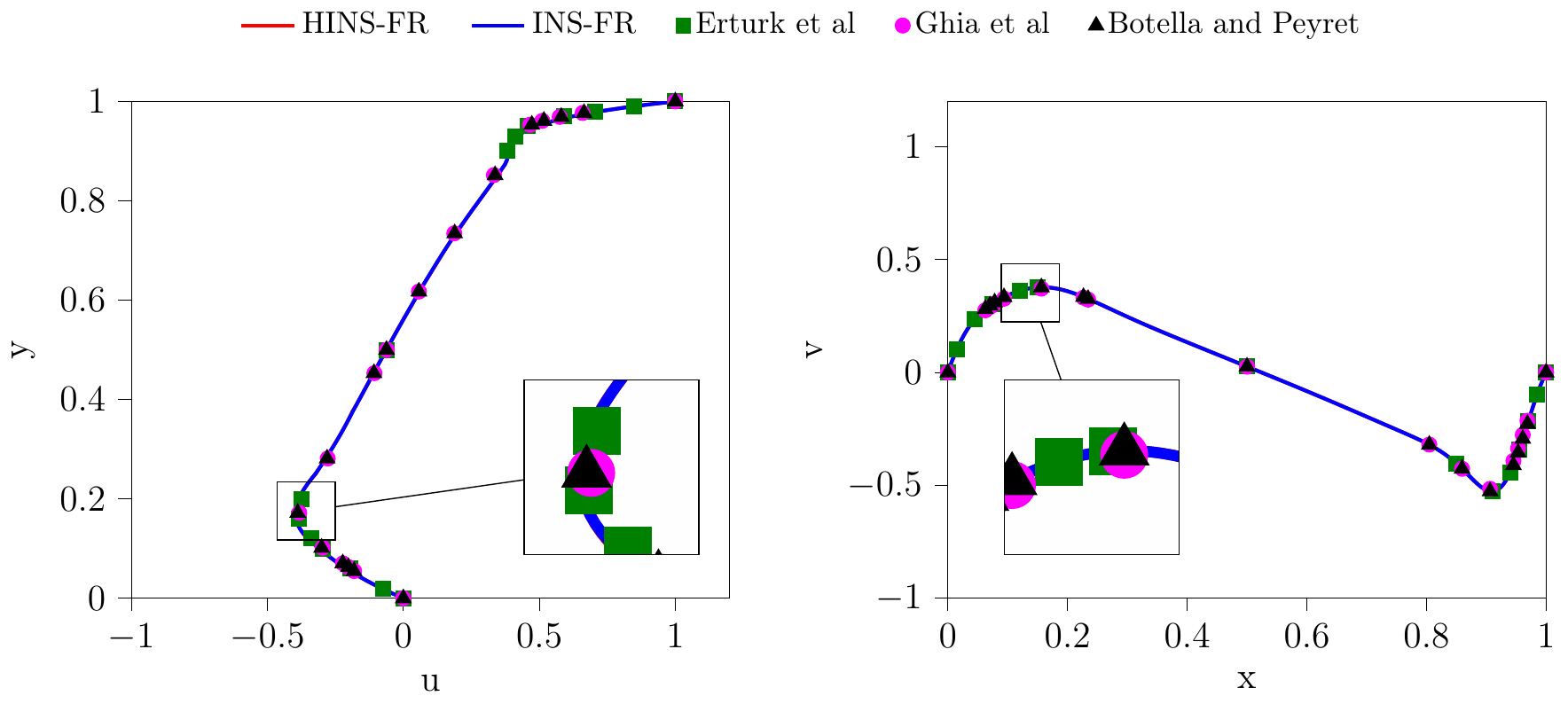}\newline
    }\\
    \resizebox{\columnwidth}{!}{%
    \includegraphics{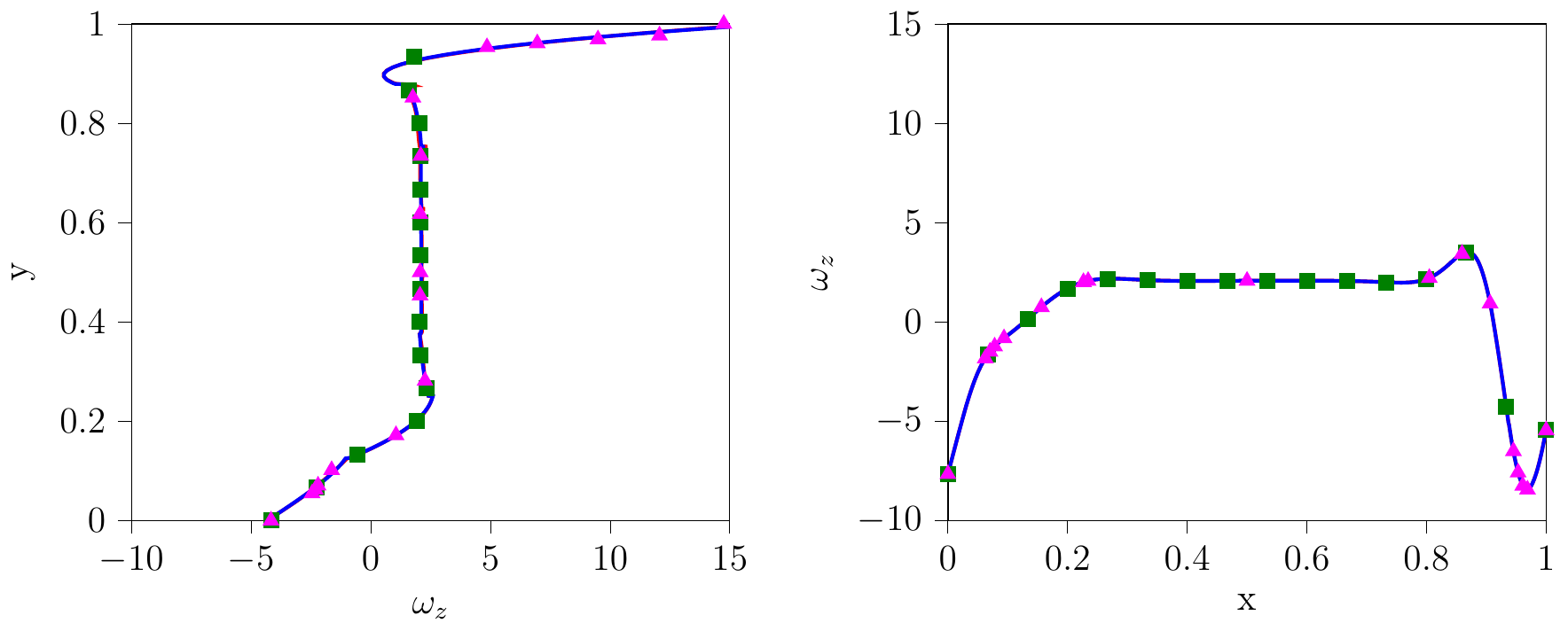}
    }
    \caption{Comparison between computed results and published literature}
    \label{figure:drivencavity_literature_compare}
\end{figure}

The problem of the flow inside a driven lid cavity is examined in this section. In this problem, the flow inside the cavity is driven by the movement of one or more walls. 
This results in a complex vortex pattern with separation and re-attachment on the cavity walls depending on the Reynolds Number as shown in \Cref{figure:drivencavity_pattern}.
The Reynolds Number $Re = {{U_{Lid} L} \over \nu}$ is set by choosing a suitable viscosity value while maintaining a unit lid velocity.
No slip wall boundary conditions are applied everywhere on the boundaries. 
% \subsubsection{Compare with Literature}
% Add three figures one for each Re [100, 1000], compare with Erturk, Botella and Ghia for velocity
% May need to revisit and re-compute at a
% much coarser mesh say 5x5 or 9x9
In \Cref{figure:drivencavity_literature_compare}, velocity and vorticity from both solvers (INS-FR and HINS-FR) are compared to reference results at the cavity mid-lines from \citeauthor{ghia1982high}\cite{ghia1982high}, \citeauthor{erturk2005numerical}\cite{erturk2005numerical} and \citeauthor{botella1998benchmark}\cite{botella1998benchmark}. 
Results were computed at $Re = 1000$ on a 8x8 uniform quadrilateral mesh with $m=3$. The figure shows that both methods produce excellent agreement with published literature with hardly any visible differences in both the velocity and vorticity distributions.
% Add three figures one for each Re [1000], compare with Erturk, Botella for vorticity

% \subsubsection{Pseudo-Stability vs Mesh-size and Reynolds Number}
% psuedo stability with mesh size
% psuedo stability with nu\cite{}

\begin{figure}[htbp!]
  \centering
  \resizebox{\columnwidth}{!}{%
  \includegraphics{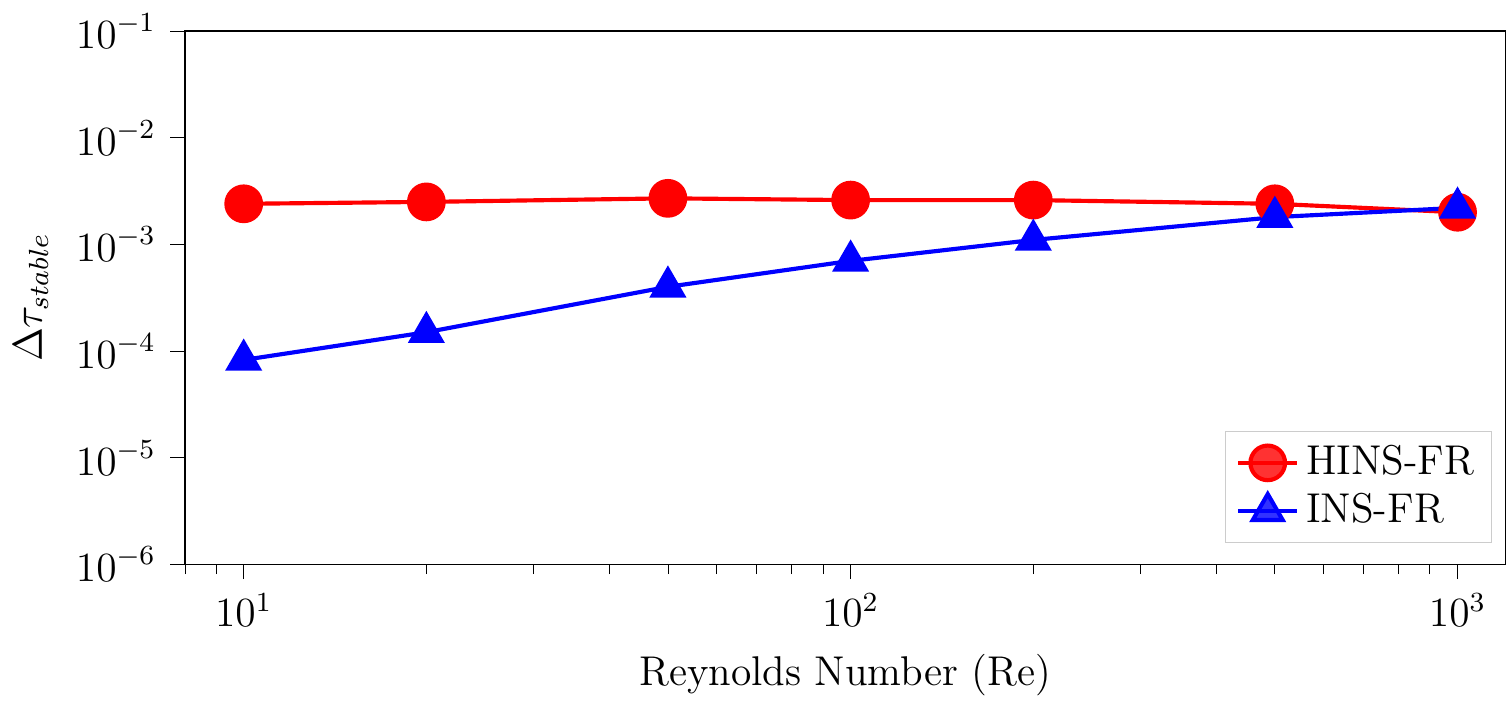}\newline
  }
  \caption{Effect of Reynolds number on the maximum pseudo-time step}
  \label{figure:drivencavity_stability_nu}
\end{figure}

\begin{figure}[htbp!]
  \centering
  \resizebox{\columnwidth}{!}{%
  \includegraphics{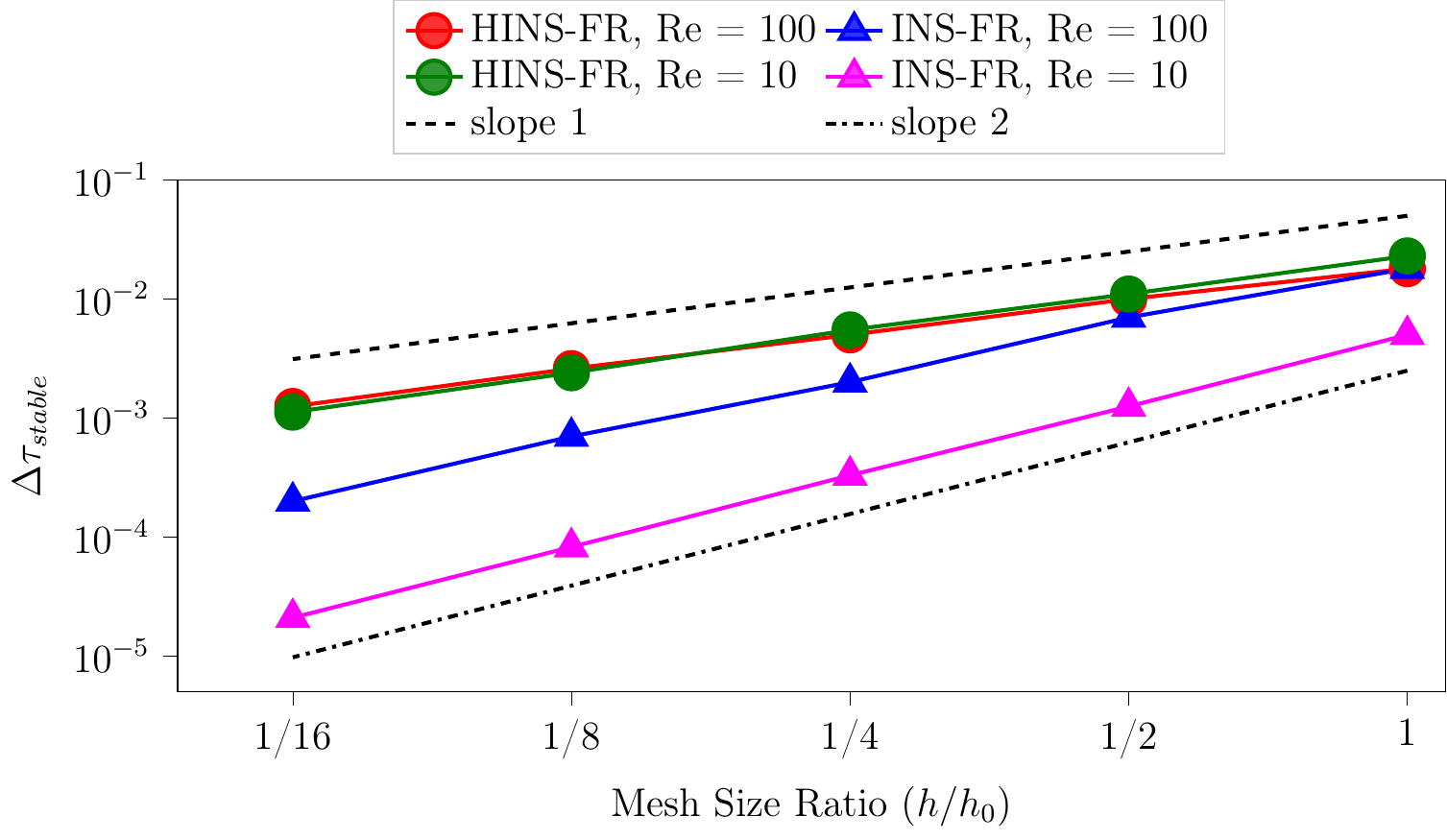}\newline
  }
  \caption{Effect of mesh resolution on the maximum pseudo-time step $\Delta\tau$ at Re = 10 and Re = 100. $h_0$ is the mesh size on the coarsest mesh}
  \label{figure:drivencavity_stability_nu_h}
\end{figure}

The effect of mesh resolution and Reynolds Number on the numerical stability of both methods is examined here.
A series of simulations are carried out for Reynolds numbers between 10 and 1000 and the maximum allowable pseudo-time step $\Delta \tau_{stable}$ is obtained by trial and error. \Cref{figure:drivencavity_stability_nu} shows that the HINS-FR solver is almost unaffected by the change in Reynolds number while the INS-FR formulation seems significantly affected by it.
%%%%%%%%%%%%%%%%%%%%%%%%%%%%%%%%%%%%%%%%%%%
Additional cases with $Re=10$ and $Re=1000$ are carried out using five different mesh resolutions in \Cref{figure:drivencavity_stability_nu_h}. 
It can be observed that $\Delta \tau_{stable}$ of HINS-FR solver  decreases linearly with mesh size while it decreases quadratically for the INS-FR solver.
The INS-FR solver  is restricted by the parabolic and hyperbolic CFL limit such that  
\[\Delta \tau < \min( {CFL}_{hyp} {h \over \max(|\lambda_{e_i}|)}, {CFL}_{par} {h^2 \over 2\nu})\]
where $\lambda_{e_i}$ is the local spectral radius of the inviscid flux jacobian. For diffusion dominated problem, the parabolic criterion poses a severe restriction on the pseudo-time step size. 
Such a restriction doesn't exist for HINS-FR where the system of equations is first order hyperbolic thus only limited by the hyperbolic CFL criterion 
\[\Delta \tau < {CFL}_{hyp} {h \over \max(|\lambda_{t_i}|)}\] 
where $\lambda_{t_i}$ is the local spectral radius of the total flux jacobian. 
% This can be confirmed by examining the slopes of $\Delta \tau_{stable}$ with mesh size which is proportional to $h$ and $h^2$ in the cases of HINS-FR and INS-FR respectively. 

% This can be rather restricting even at high Reynolds number when $\mu = \mathcal{O}(h^2) $ which is a common situation in Direct Numerical Simulation (DNS) and Large Eddy Simulation (LES).
% modify this bad parah

% \bibliographystyle{model1-num-names}
% \bibliography{references.bib}

% \end{document}

% % \newpage\mathcal{J} 
%% main textP
\subsection{Flow over Sphere}
\begin{figure}[htbp]
  \centering
  \includegraphics[width=0.48\linewidth]{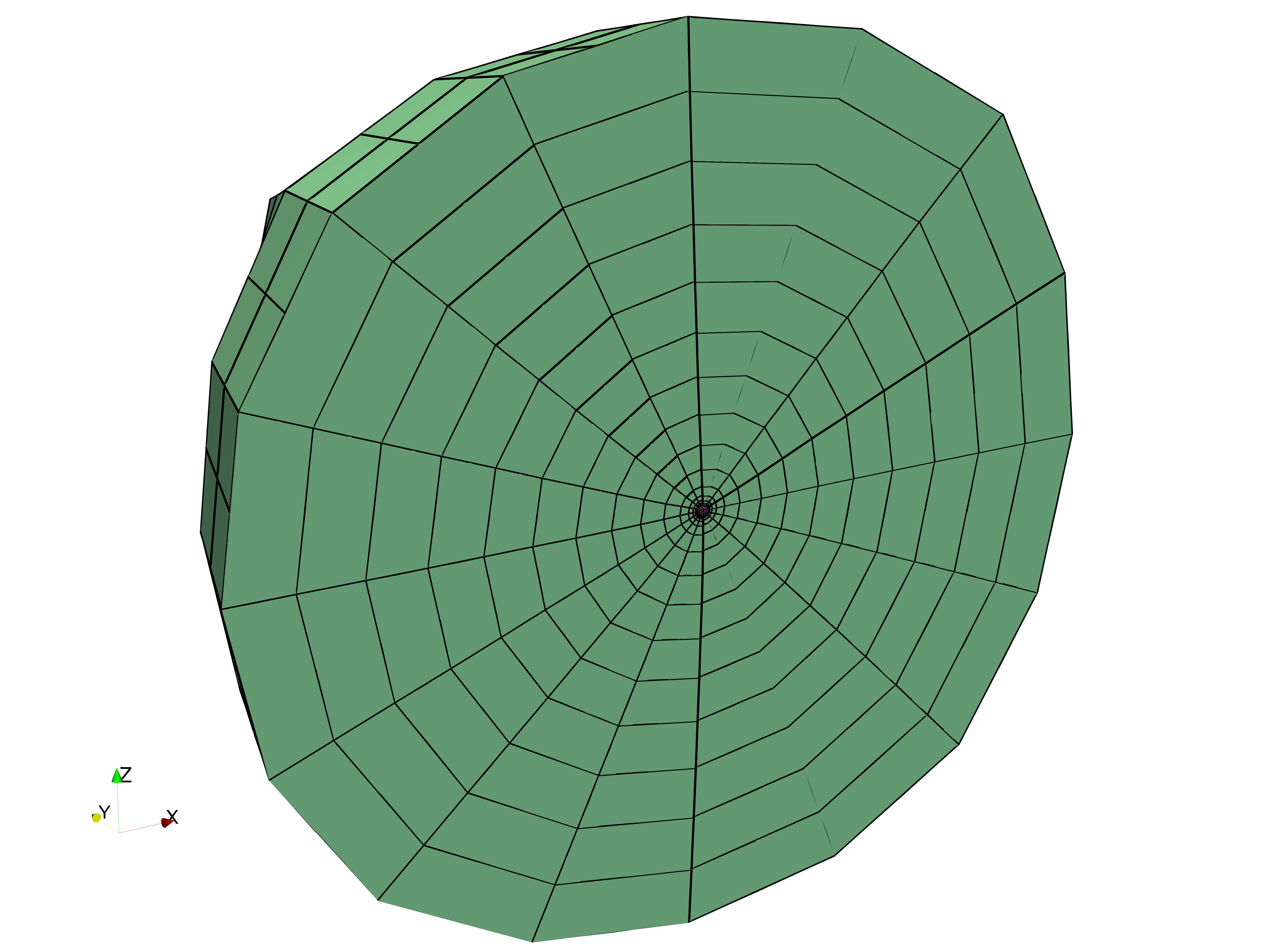}
  \includegraphics[width=0.48\linewidth]{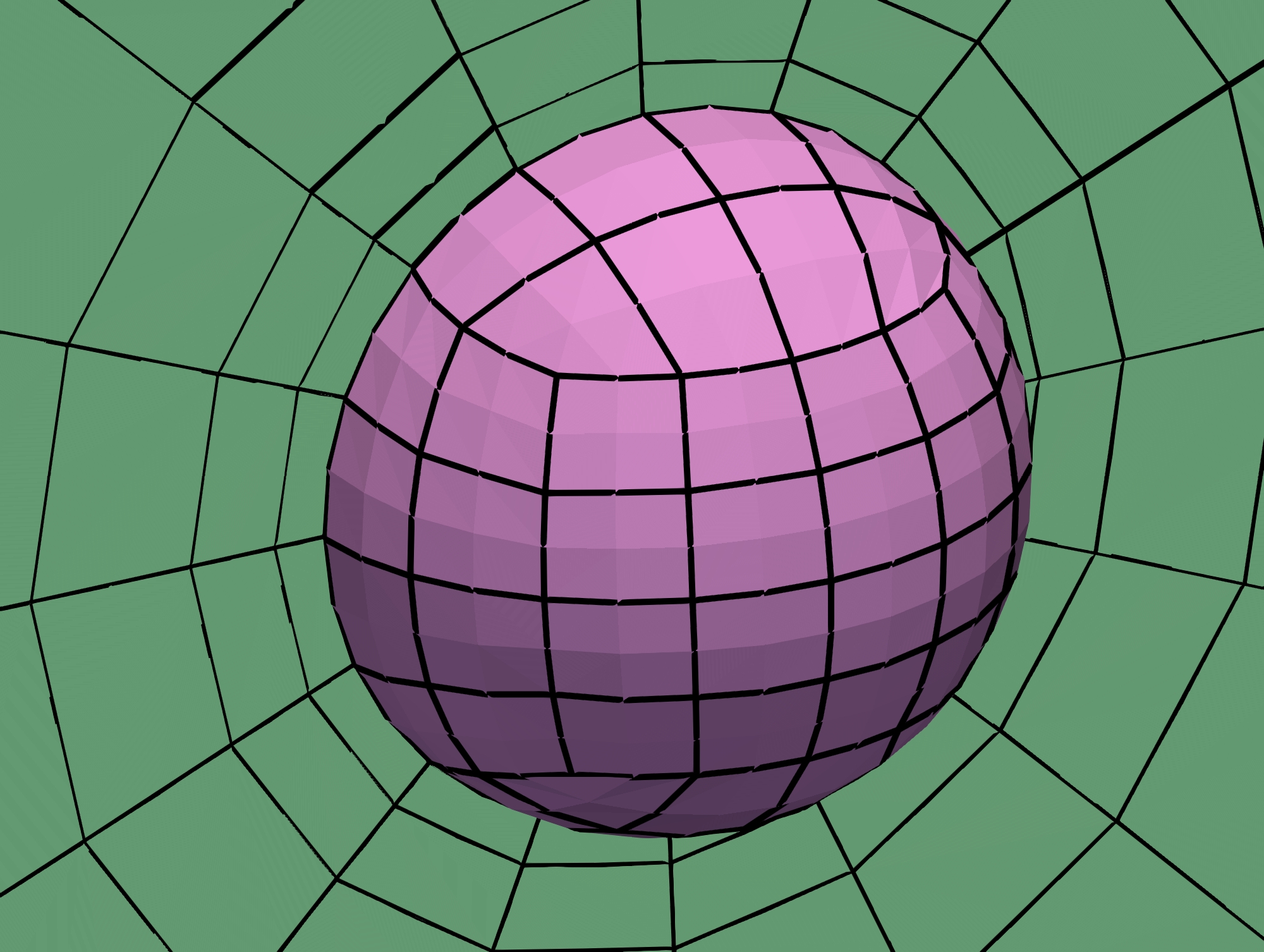}
  \caption{3D hexahedral mesh for the flow over a sphere at Re = 20. Mesh count is 1440 elements}
  \label{figure:sphere_mesh_coarse_re20}
\end{figure}

The steady laminar flow over a sphere is simulated here to further verify the computational accuracy and evaluate the efficiency of the developed HINS-FR method. The Reynolds number is set to $Re = 20$. The simulations were performed using polynomial orders $m=~1 ~\text{through}~ 5$ on a very coarse O-type mesh, shown in \Cref{figure:sphere_mesh_coarse_re20}. The mesh contained only 1440 hexahedra with the far-field at 50 times the sphere diameter. 
In order to accurately capture the curvature of the sphere, second-order curved elements are used. 

In this test case, the results of both incompressible FR solvers are compared with published data computed using the high-order DG method\cite{CRIVELLINI2013442} on a similarly coarse mesh as well as experiments.
 
\begin{table}[htbp!]
  \caption{Comparison of computed drag coefficient with relevant numerical results and experiments in literature at Re = 20}
  \label{table:sphere_re20}
  \begin{tabular}{@{}llllll@{}}
    \toprule
    \multirow{2}{*}{}             & \multicolumn{5}{c}{Coefficient of Drag ($C_D$)}    \\ \cmidrule(l){2-6} 
                                  & $m=1$     & $m=2 $    & $m=3$     & $m=4$     & $m=5$     \\ \midrule
   HINS-FR                       & 2.7253 & 2.7248 & 2.7195 & 2.7192 & 2.7193 \\
    INS-FR                        & 2.5486 & 2.7828 & 2.718 & 2.7196 & 2.7197 \\
    \midrule
    Discontinuous Galerkin (DG) \cite{CRIVELLINI2013442}   & 2.882   & 2.749   & 2.782   & 2.721   & 2.719   \\
    \midrule
    \begin{tabular}[c]{@{}l@{}}Numerical Computation: \\ \citeauthor{tabata1998precise}\cite{tabata1998precise}\end{tabular} & \multicolumn{5}{c}{2.724} \\
    Exp. Data: Schlichting \cite{schlichting2016boundary}          & \multicolumn{5}{c}{2.79}                        \\ \bottomrule
    \end{tabular}
  \end{table}
The tabulated results given in \Cref{table:sphere_re20} show that the developed HINS-FR solver achieves excellent agreement even with a first order polynomial reconstruction.
For INS-FR, a third-order polynomial reconstruction is needed to achieve similar accuracy.
On the other hand, fourth order polynomial reconstruction is required for the DG method.
For both HINS-FR and INS-FR solvers, the computed drag coefficient becomes independent of the reconstruction order for $m\geq3$. 

\begin{figure}[htbp]
  \centering
  \includegraphics[width=0.4\linewidth]{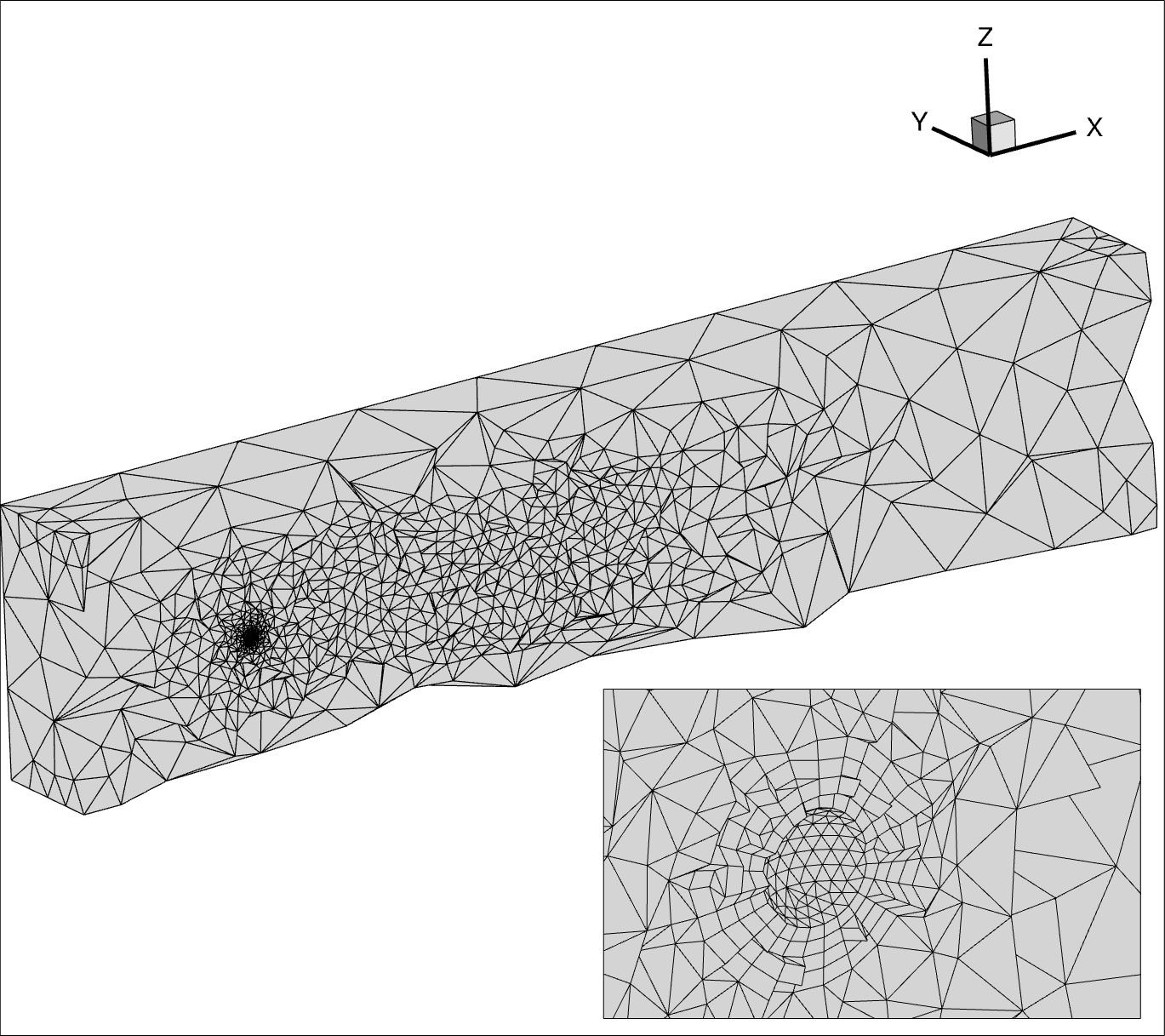}
  \qquad
  \includegraphics[width=0.4\linewidth]{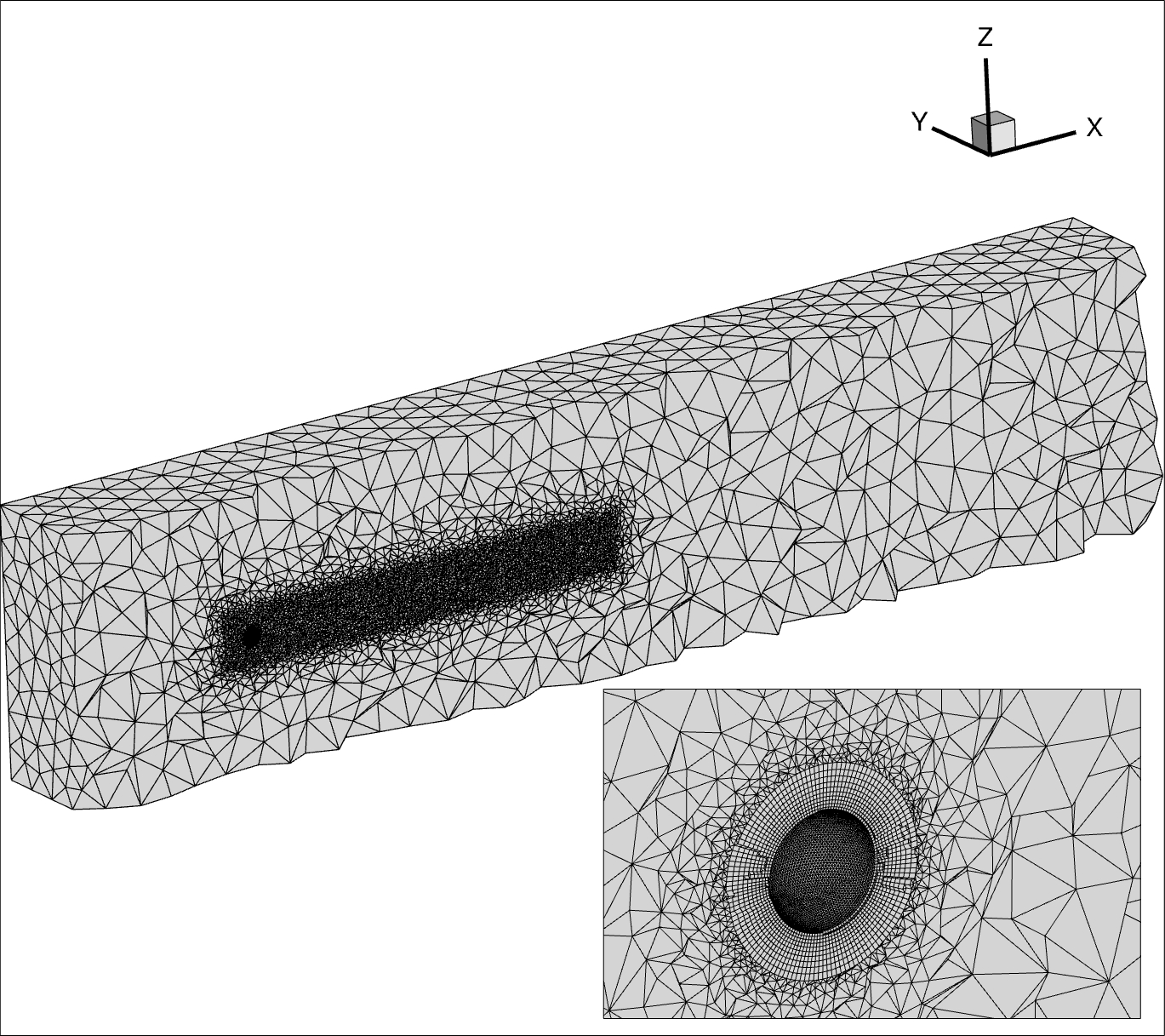}
  \caption{3D mixed tetrahedral-prism mesh for the flow over sphere at Re = 100. Base mesh: 35K cells (left). Fine mesh: 750K cells (right)}
  \label{figure:sphere_mesh_coarse_fine}
\end{figure}

  %%%%%%%%%%%%%%%%%%%%%%%%%%%%%%%%%%%%%%%%%%%%%%%
An additional case is performed for $Re = 100$ where mixed meshes are employed as shown in \Cref{figure:sphere_mesh_coarse_fine}. 
This case is also a steady case showing an axisymmetric wake field. 
The case serves to demonstrate the accuracy and efficiency of the developed for mixed mesh topologies.  It also allows sufficient quantitative comparison with other published results.

Two meshes are constructed, with prism layers near the surface of the sphere and tetrahedra through the domain, and denoted base mesh and fine mesh. 
A refinement box is set downstream of the sphere to properly capture the vortex structure due to separation and  wake dynamics. The base mesh used in the simulation is made up of nearly 35K elements. This mesh is used to compute the solution at different reconstruction orders namely, $m=1,~2,~3,~ \text{and}~ 4$.
The solution is also computed on a more refined mesh with 750K elements and it is used to verify our previously obtained results and serve as benchmark results for future simulations. Consequently, it is only computed for third order reconstruction.
The pressure and velocity field are initialized using freestream values and the simulation is performed until the residual of all variables reach steady-state with a tolerance of $10^{-9}$.

\begin{figure}[htbp]
  \centering
  \includegraphics[width=0.4\linewidth]{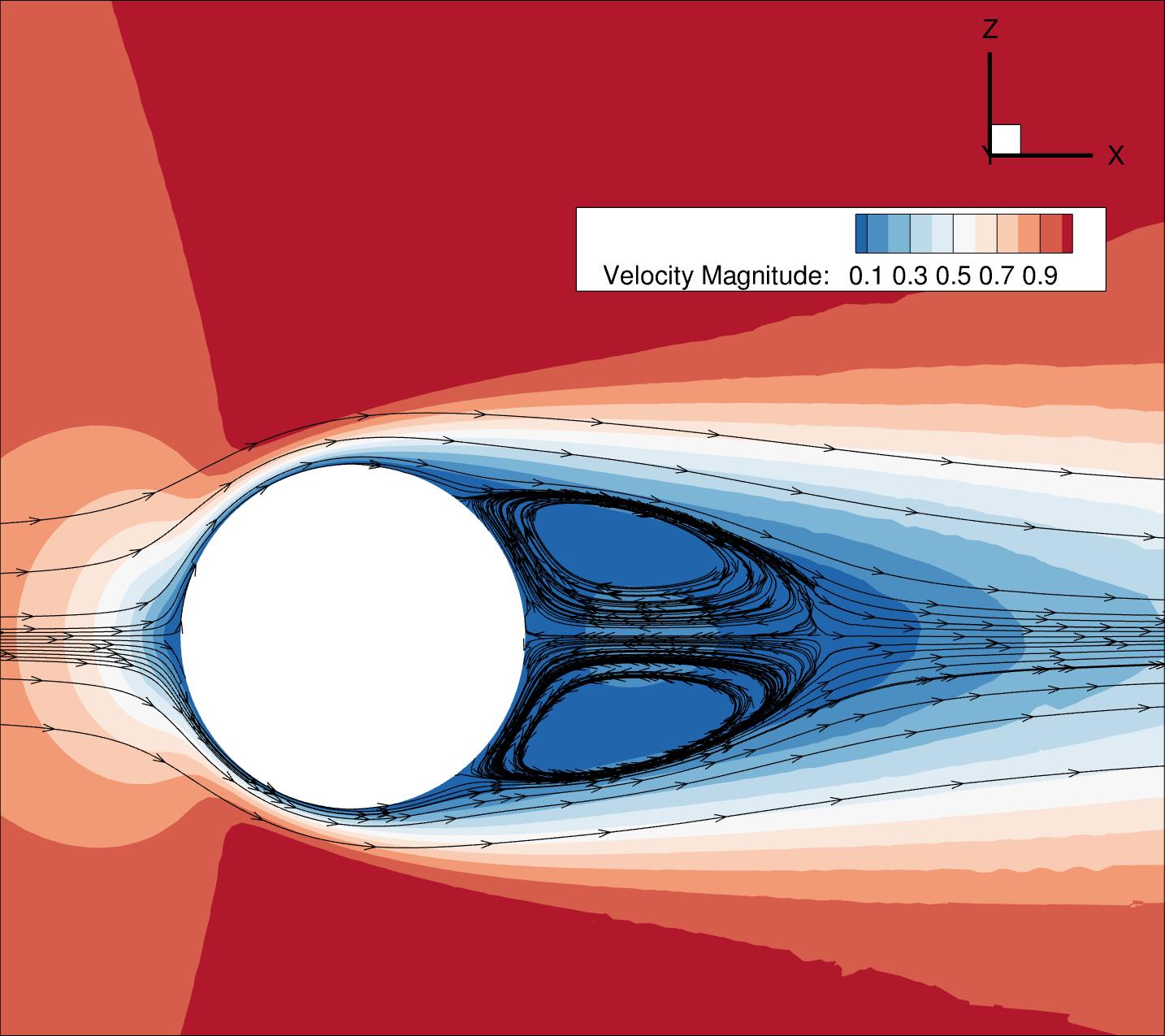}
  \qquad
  \includegraphics[width=0.4\linewidth]{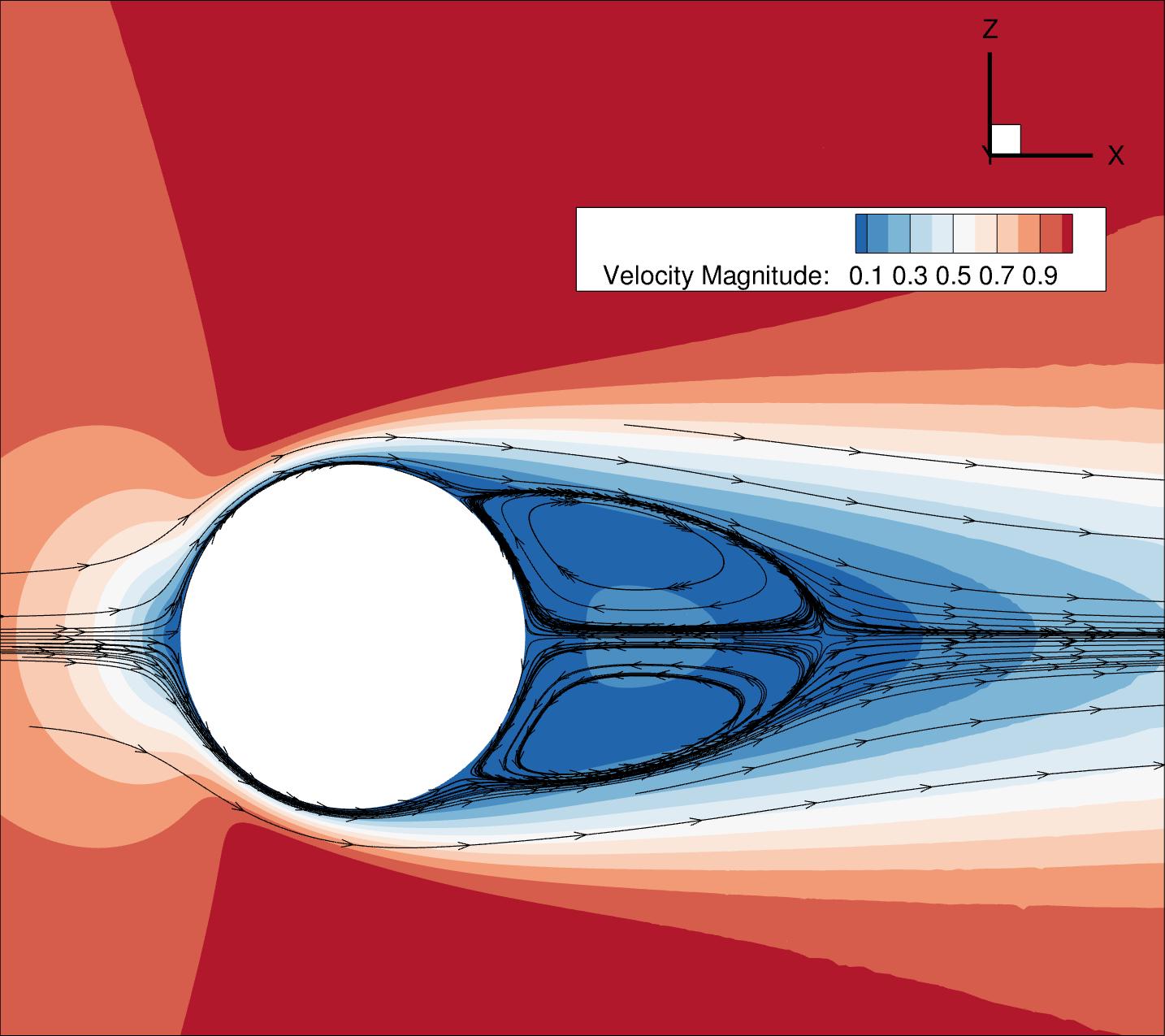}
  \caption{Velocity magnitude contours and streamlines for the flow over sphere at Re = 100 computed using the developed HINS-FR. Base mesh: 35K cells (left). Fine mesh: 750K cells (right)}
  \label{figure:sphere_sep_coarse_fine}
\end{figure}
% revise the table
\begin{table}[htpb!]
  \caption{Comparison of sphere drag coefficient and length of recirculation zone at Re = 100 with relevant numerical and experiments results from literature}
  \label{table:sphere_compare_cd_ld}
  \centering
  \begin{tabular}{@{}l|c|c@{}}
  Data                           & $C_D$& $L_r/d$ \\ \midrule
  \multicolumn{3}{l}{Present Method, Base Mesh}                                           \vspace{4pt}\\
  HINS-FR ($m =~ 1$)                &    1.01706             &    0.5826  \\
  HINS-FR ($m =~ 2$)                &    1.10419             &    0.8624  \\
  HINS-FR ($m =~ 3$)                &    1.09049                    &  0.8618    \\
  HINS-FR ($m =~ 4$)                &    1.08705             &   0.8611   \\
  INS-FR ($m =~ 1$)                 &    0.95270             &  0.6234    \\
  INS-FR ($m =~ 2$)                 &    1.13195             &  0.8774    \\
  INS-FR ($m =~ 3$)                 &    1.09640             &  0.8624    \\
  INS-FR ($m =~ 4$)                 &    1.08659             &  0.8628    \vspace{4pt}\\
  \multicolumn{3}{l}{Present Method, Fine Mesh}                                           \vspace{4pt}\\
  HINS-FR ($m =~ 3$)     &    1.08818             &  0.866    \\
  INS-FR ($m =~ 3$)      &    1.08817             &  0.865    \vspace{4pt}\\
  \multicolumn{3}{l}{Numerical Computations}                          \vspace{4pt}         \\
  HINS-FVM\cite{Hyung2020104434}  & 1.109                  & -    \\
   INS-FVM\cite{Hyung2020104434}  & 1.091                  & -    \\
   Spectral collocation method\cite{mittal1999fourier}  &   1.09                     & 0.87     \vspace{4pt}\\
  \multicolumn{3}{l}{Experiments} \vspace{4pt}\\
  \citeauthor{roos1971some}\cite{roos1971some}      &             1.08           &     \\
  \citeauthor{clift2005bubbles}\cite{clift2005bubbles}        &             1.09           &     
  \end{tabular}
  \end{table}
  % Ahn (HINS-FVM,Mesh = 1,557,038 )
In \Cref{table:sphere_compare_cd_ld}, the computed drag coefficient and length of recirculation are compared 
% between HINS-FR and INS-FR 
for different reconstruction order $m$. with results from published literature.

Both HINS-FR and INS-FR give the same value for the drag coefficient on the fine mesh. For the base mesh, the HINS-FR  consistently gives better agreement with the fine mesh result when compared to INS-FR.
Although both methods agree well with other published results, the HINS-FR gives more consistent results with increasing order. 
Additionally, when compared to the finite-volume method implementation of the hyperbolic incompressible solver, the flux reconstruction solver is able to produce comparable, if not better, results with fewer degrees of freedom.

\newpage
%% Comparing convergence history 

\begin{figure}[htbp]
  \centering
  \includegraphics[width=0.48\linewidth]{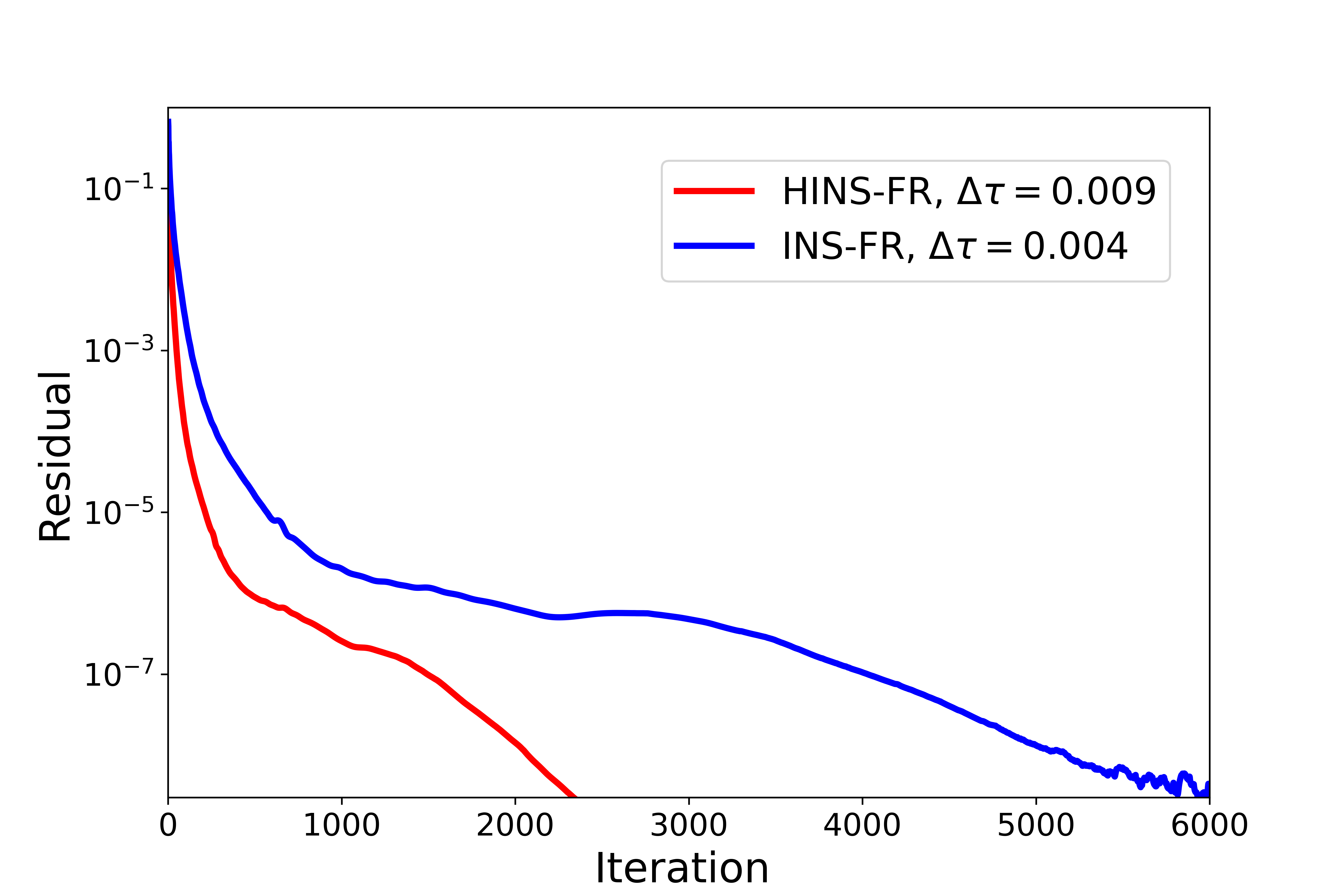}
  \includegraphics[width=0.48\linewidth]{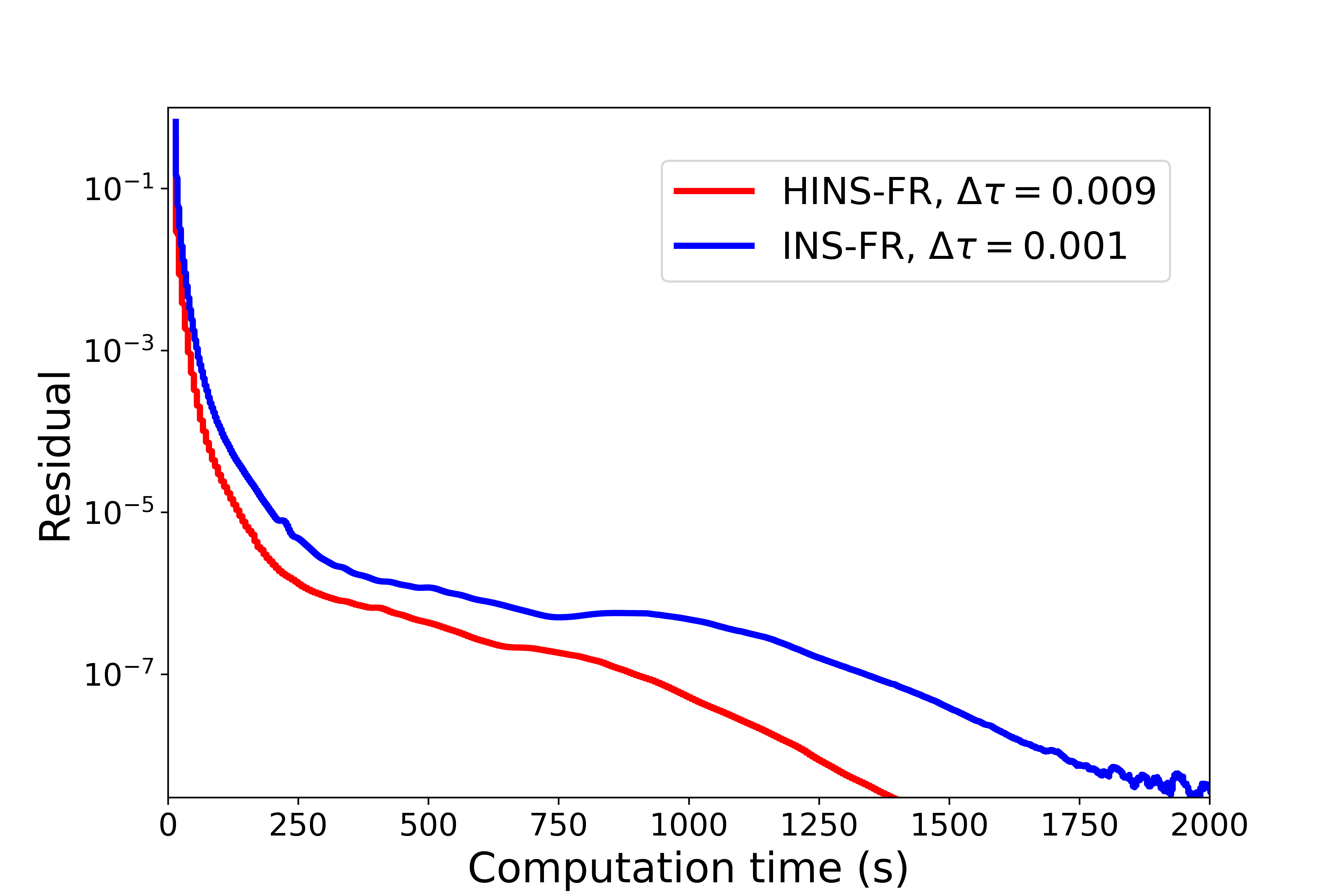}
  \caption{Convergence of the residuals of pressure for laminar flow past a sphere at Re=100 on the base mesh at $m=~3$}
  \label{figure:sphere_histroy}
\end{figure}
% The convergence of the pressure on the base mesh
We next examine the convergence performance of both solvers for the $Re =100$ case on the base mesh with the maximum stable pseudo time-step  used for each solver.
The HINS-FR requires a pseudo time-step that is 2.25 times higher than that for the INS-FR which is clearly reflected in the number of iterations to convergence shown in \Cref{figure:sphere_histroy}.
However, when the residual is plotted against the computation time, the difference shrinks drastically with the HINS-FR still leading in terms of performance with a ratio of 1.45. 
This indicates that the computation cost per iteration is higher in the case of HINS-FR solver. This is understandable considering that the number of variables in the case of the  hyperbolic incompressible Navier-Stokes formulation in 3D is 13 as compared to 4 for the conventional formulation.

%%%%%%%%%%%%%%%%%%%% Performance %%%%%%%%%%%%%%%%%%%%%%
 In order to compare the scalability and efficiency of the HINS-FR and INS-FR solvers, a strong and weak scaling study was also carried out. A mesh with 73,372 hexahedral elements was used for this study. The reconstruction polynomial order was set to $m = 3$. 
A 4 level P-multigrid Runge-Kutta-Vermeire smoother cycle 1-1-1-2-1-1-2 was found to give a good balance of residual reduction vs computational cost. 
Double precision was used for all computations considered here. 
\begin{figure}[htbp!]
  \centering
  \includegraphics[width=0.45\linewidth]{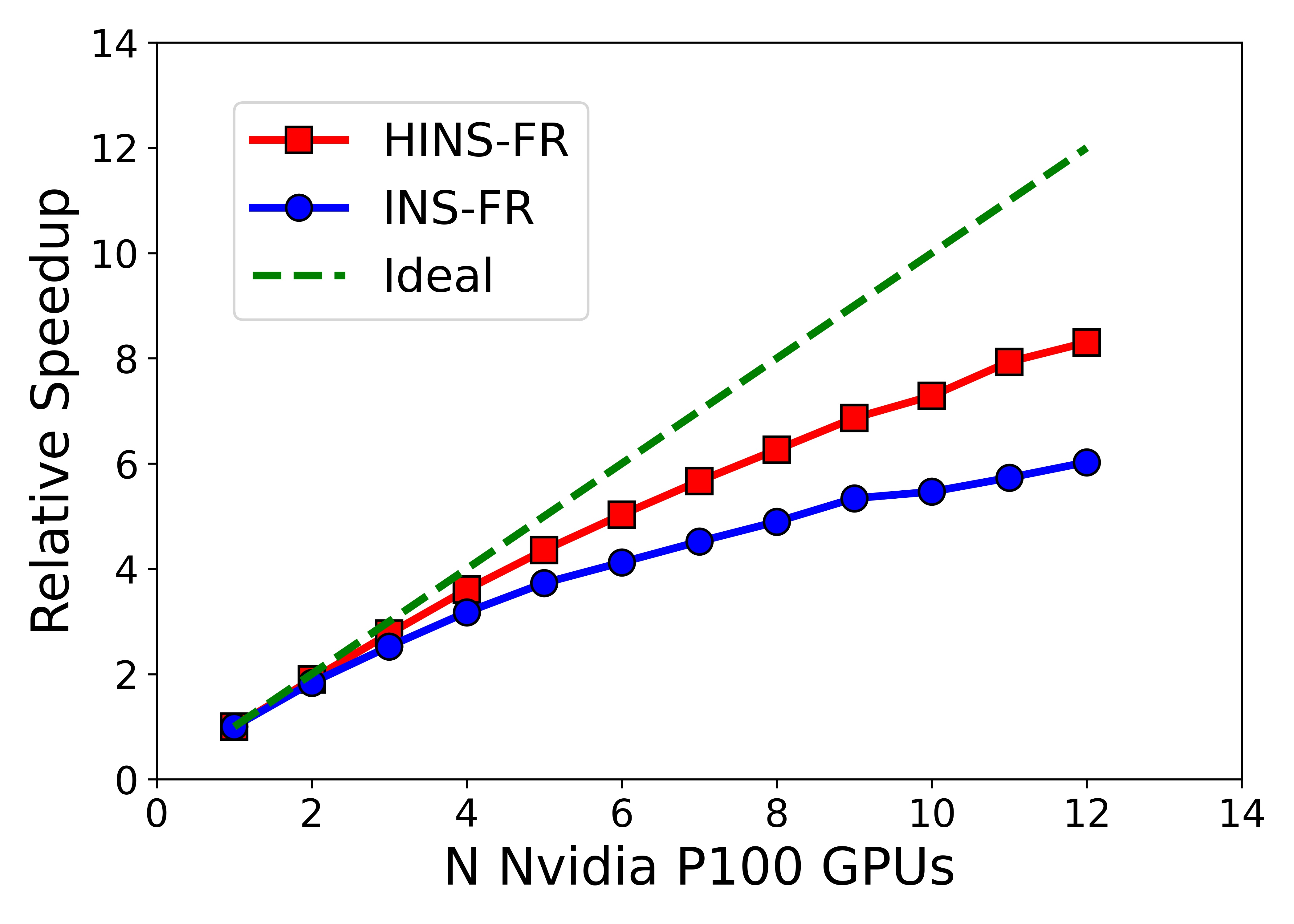}
  \includegraphics[width=0.45\linewidth]{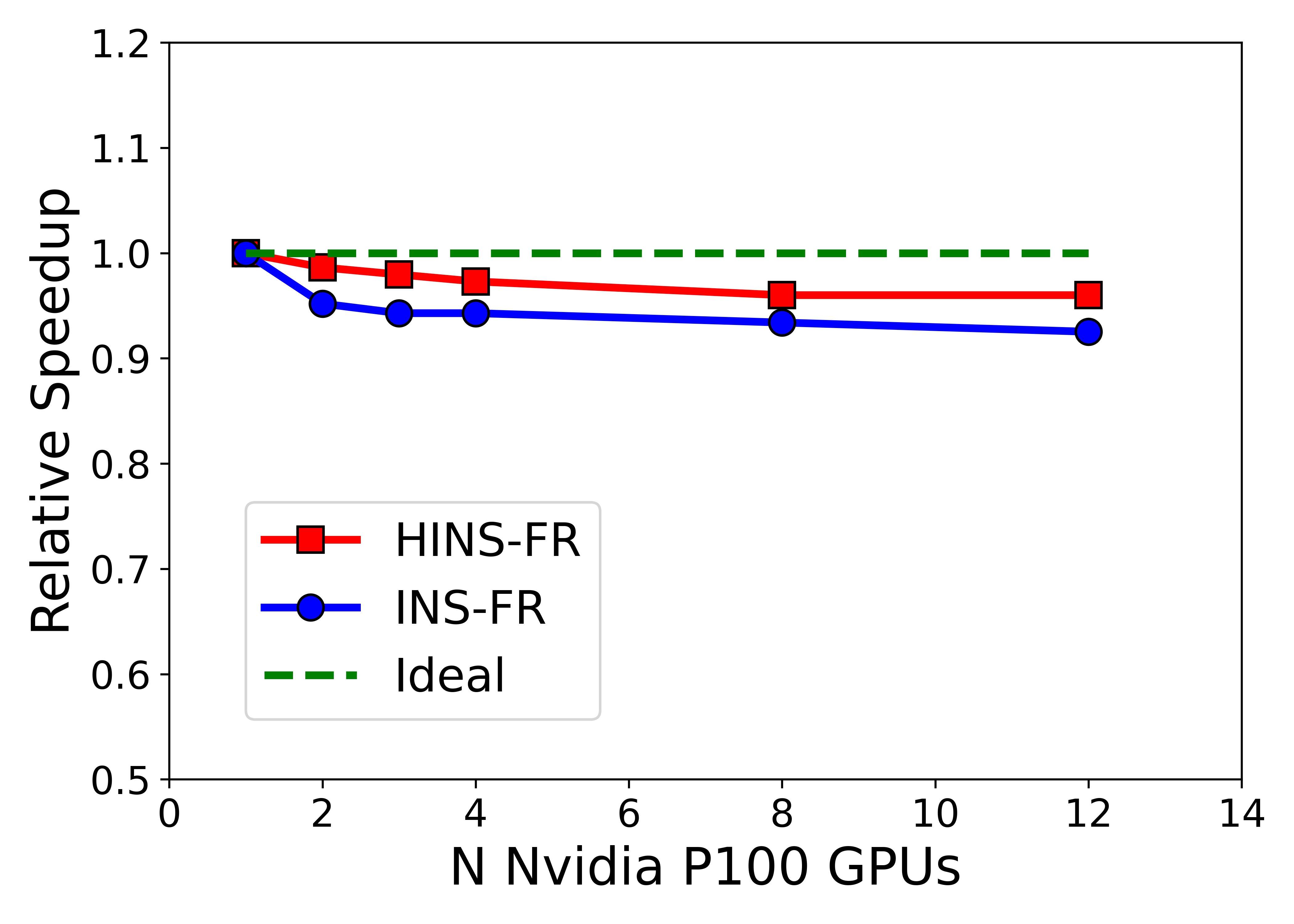}
  \caption{Strong Scaling of the incompressible flow past a sphere
   on NVIDIA Tesla P100 GPUs with p-Multigrid for HINS-FR and INS-FR solvers}
  \label{figure:performance_scaling_strong_weak}
\end{figure}

\Cref{figure:performance_scaling_strong_weak} shows the strong and weak scaling for both solvers from 1 through 12 NVIDIA P100 GPUs.
Strong scaling in both cases is almost linear up to 6 GPUs after which both solvers start to branch out.  
The INS-FR solver experiences a quicker decline due to extra communication required by the LDG procedure. 
Regarding weak scaling, both solvers are able to maintain relatively good performance with an efficiency higher than $90\% $ with the INS-FR solver slightly under-performing when compared to HINS-FR solver.
% % \newpage
\section{Conclusion}\label{sec:conc}
A high order hyperbolic incompressible Navier-Stokes solver has been developed using the flux reconstruction approach.  
The developed solver has been implemented in the cross-platform PyFR framework using the hyperbolic formulation of the artificial compressibility method.  
Significant reduction in the absolute error of the field variables and the gradient of the velocity has been demonstrated .  
Additionally, it has been shown that equal orders of accuracy can be obtained for both the field variables and velocity gradients. 
Numerical results suggests that the improvement in the order of accuracy of the velocity gradient lead to a matching improvement of the pressure order of accuracy. 
Analysis shows that the time-step requirements are  significantly relaxed when using the hyperbolic solver.  
This is because the parabolic CFL criterion is $\mathcal{O} (h^2)$ while the hyperbolic CFL criterion is $\mathcal{O} (h)$. 
This leads to significant convergence speed-ups especially for diffusion dominated problems where the parabolic restriction can be quite severe. 
The strong scaling performance of the developed HINS-FR solver has been shown to be superior to the existing INS-FR solver due to the extra communication required for the computation of the viscous fluxes. 
In conclusion, the hyperbolic method is appealing owing to its accuracy, stability and efficiency in solving diffusion dominated problems. 

While the currently developed solver can be used for unsteady problems via dual-time marching, there is more to be desired regarding its performance. The current code implementation requires some optimization to better handle the sparsity of the flux vector. This will lead to significant improvements in memory foot-print and computational performance. 
\section*{Acknowledgment}
This work was supported in part by JSPS KAKENHI Grant Number JP19H02363.
The numerical calculations were carried out on the TSUBAME 3.0 supercomputer at Tokyo Institute of Technology.

\bibliographystyle{model1-num-names}
\bibliography{references.bib, references_hyperbolic.bib}

\end{document}